\documentclass[aps,pre,twocolumn,superscriptaddress,showpacs]{revtex4}
\usepackage{epsfig}
\usepackage{changebar}
\usepackage{graphicx}
\usepackage{epsfig}
\usepackage{subfigure}
\usepackage{amsmath}
\usepackage{rotating}
\usepackage{amsfonts}
\usepackage{amssymb,amsmath} 
\psfigdriver{dvips}
\begin{document}
\title{Thermal shifts and intermittent  linear response of aging systems }  
\author{Paolo Sibani and Simon Christiansen }
\email[]{paolo.sibani@ifk.sdu.dk}
\affiliation{Institut for Fysik og Kemi,  SDU, DK5230 Odense M, Denmark} 
\begin{abstract}
At  time $t$ after an initial   quench,   an aging system responds to 
a   perturbation  turned  on
at time $ t_{\rm w} < t$  in a way mainly depending  on the   number  
of  intermittent energy fluctuations, so-called
quakes, which fall  within the observation interval $(t_{\rm w},t]$ 
[Sibani et al. Phys. Rev. B, 74, 224407 and Eur. J. of Physics B, 58,483-491, 2007].
The temporal distribution of the quakes  implies a  functional dependence of the    average response  on the ratio
 $t/t_{\rm w}$. Further insight is obtained imposing small  temperature steps,  so-called  $T$-shifts.  
The  average response  as  a function of  $t/t_{\rm w,eff}$, where  $t_{\rm w,eff}$  is
the effective age, is similar to the response of a system aged isothermally at 
the final temperature.    
Using an Ising model  with plaquette interactions,    the   applicability  of analytic
formulae for the average isothermal  magnetization is confirmed.  
The  $T$-shifted aging behavior of the model is described using effective ages. 
Large positive shifts nearly reset the effective  age. 
Negative $T$-shifts  offer a  more detailed probe of  the dynamics 
. Assuming the marginal stability of the `current' attractor  against  thermal noise fluctuations,  
the  scaling  form $t_{\rm w,eff} = t_{\rm w}^x$, and  the  dependence of the  exponent 
 $x$  on the aging temperatures  before and after  the shift are theoretically available.   
 The predicted form of $x$ has no adjustable parameters. Both the algebraic scaling of the
 effective age and the form of the exponent agree with the data. The simulations     thus  confirm 
the crucial r\^{o}le of marginal stability in glassy relaxation.
\end{abstract} 
 
\maketitle
\section{Introduction} 
Rejuvenation, memory    and other intriguing 
aspects of  out-of-equilibrium thermal relaxation, a process
widely  known as  aging,   continue to 
 attract  experimental and theoretical interest  
~\cite{Lundgren86,Schultze91,Sibani91,Hoffmann97,Jonason98,Komori99,Komori00a,Komori00b,Normand00,Utz00,Bouchaud01,Nicodemi01,Bonn02,Berthier02,Takayama02,Jensen02,Sibani04a}.

One starting point for  theoretical analyses is the 
real space morphology of spin-glasses with short-range interactions, where  
thermalized domains  form, with a    linear size growing in time as 
$R(t)  \propto  t^\lambda$ ~\cite{Rieger93,Komori99,Komori00a,Komori00b}. The   domain size  
corresponds to the thermal correlation length and can be   extracted
from numerical simulation data. 
Furthermore, the    temperature dependence of  the exponent $\lambda$     can be rationalized 
by assuming that  the  energy  barrier controlling  domain growth increases logarithmically 
in time~\cite{Komori00b,Rieger93}. 
The same  logarithmic time dependence  characterizes    hierarchical models,   
   which   account for  e.g.
   memory effects  using    configuration space, or 'landscape' 
 properties as starting point~\cite{Vincent91,Hoffmann97,Dupuis02}. 
A third  theoretical  approach focuses on the relationship between linear response and  
autocorrelation functions.      
Numerical~\cite{Andersson92,Sibani07}  
and experimental~\cite{Svedlindh89,Herisson02,Buisson03} results
show that the   Fluctuation Dissipation Theorem (FDT)  
is applicable for  observation times  $t_{\rm obs} < t_{\rm w}$, where $t_{\rm w}$
is the time at which the  perturbation is applied.   While the FDT is broken  for larger $t_{\rm obs}$,  
 the  proportionality between 
conjugate response and correlation functions may  be  restored asymptotically for    large $t_{\rm w}$ and $t_{\rm obs}$,
whence a general  description flows in terms of 
effective temperatures~\cite{Cugliandolo97,Castillo03,Buisson03,Calabrese05}.  
Last but not least, fluctuation spectra~\cite{Crisanti04,Sibani05,Sibani06a,Sibani06b,Sibani07,Christiansen07}   
indicate that  thermally activated aging processes,   linear response functions included,
are controlled by   intermittent transitions  between   meta-stable attractors
 which irreversibly  release the  excess energy trapped in the initial configuration.
These events, dubbed   quakes, have  statistical properties suggesting 
 that they might be  triggered by  thermal  fluctuations of 
 record  magnitude~\cite{Sibani93a,Sibani93}.
   
Based on the idea that metastable attractors of glassy systems
are marginally stable, and that  the transitions between them are hence induced by
record sized thermal fluctuations, a  `record dynamics' scenario~\cite{Sibani03}     provides  approximate 
 analytical  formulae for several physical quantities.
 The derivations  ignore  quasi-equilibrium  properties, 
but are  sufficiently general to   support   the  universal character of off-equilibrium  aging  
 phenomena
 ~\cite{Cipelletti00,Josserand00,Hannemann05, Anderson04,Oliveira05,Sibani93,Rieger93}. 
 To  buttress  a   comprehensive
description of non-equilibrium aging based on    record dynamics, this   paper  
discusses numerical simulations  of an Ising model with 
plaquette interactions, which has a   glassy dynamical regime at low temperatures. 
The work extends   numerical investigations~\cite{Sibani06b,Christiansen07} of  the
same model,  and   experimental and numerical studies of linear response in  
spin-glasses~\cite{Sibani06a,Sibani07}.  

Aging under   isothermal conditions is considered first,  and analytic formulae  
   for the average magnetization~\cite{Sibani06a,Sibani07} are verified.
  Secondly,  the effect of  `$T$-steps'  or  `$T$-shifts',   small 
  temperature steps of either sign   applied  concurrently with the  magnetic field
  is analyzed in terms of   \emph{effective ages}. Replacing 
  the  age $t_{\rm w}$ as a scaling parameter,   the effective age makes the  response look similar 
  to  the    isothermal response  at the final temperature.  
   Effective ages give
  an excellent   parameterization of the   average response for negative shifts, 
   and  a reasonable  one  for positive shifts.
Fluctuation spectra cannot be fully described by an effective age.
     
For  negative T-shifts,  an  algebraic 
relation between  effective and true age exists~\cite{Sibani04a}, and     
 the  analytical dependence of the corresponding exponent   
on the temperatures before and after the step  has    no adjustable
parameters.
Both  the algebraic scaling and the for of the exponent  are confirmed by the   simulations. 
The final discussion emphasizes the  ramified  connections of the present  approach
to  other descriptions of aging dynamics.
\begin{figure*}[t!]
\centering
\mbox{
\subfigure[]{\epsfig{figure=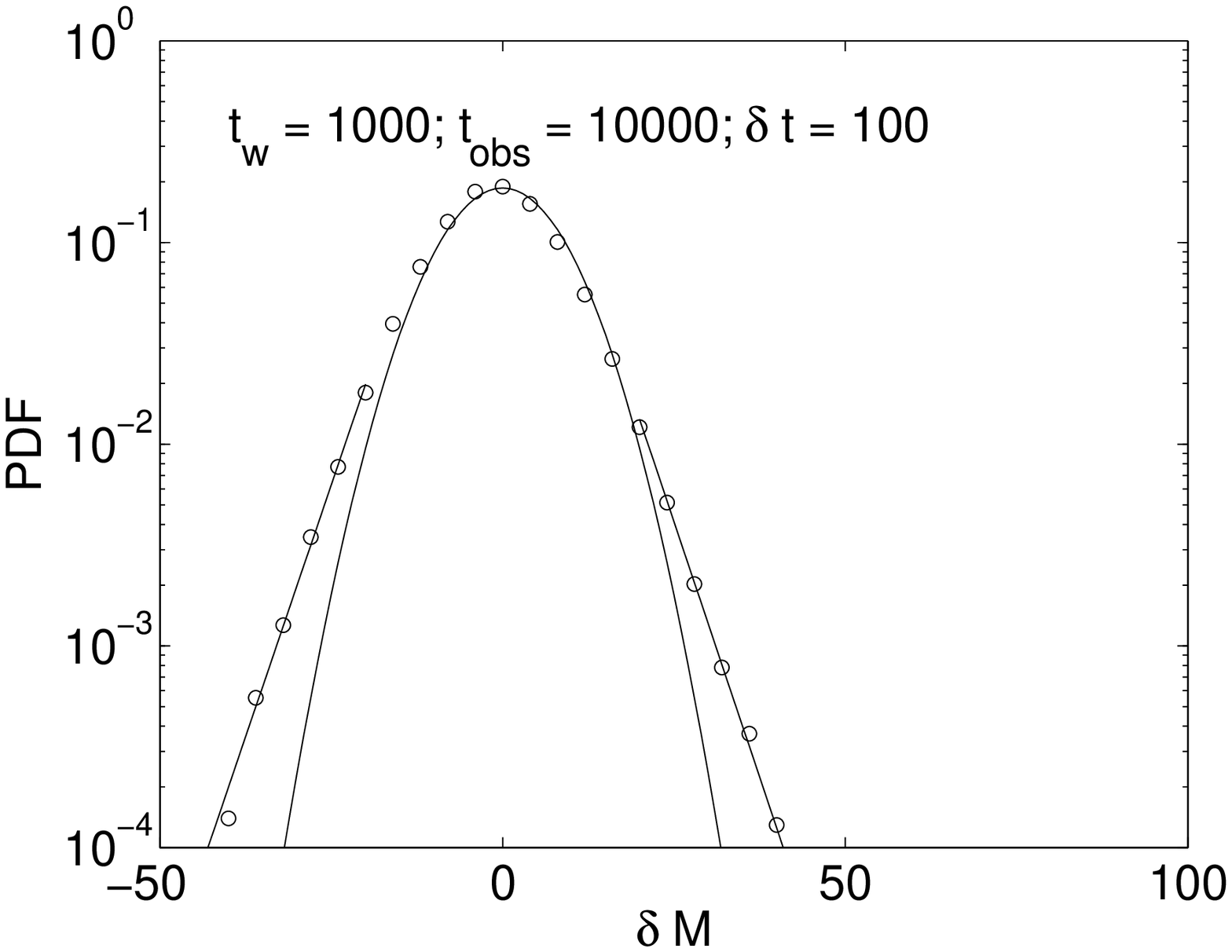,width=.47\textwidth}} \hspace{-.7cm} 
\subfigure[]{\epsfig{figure=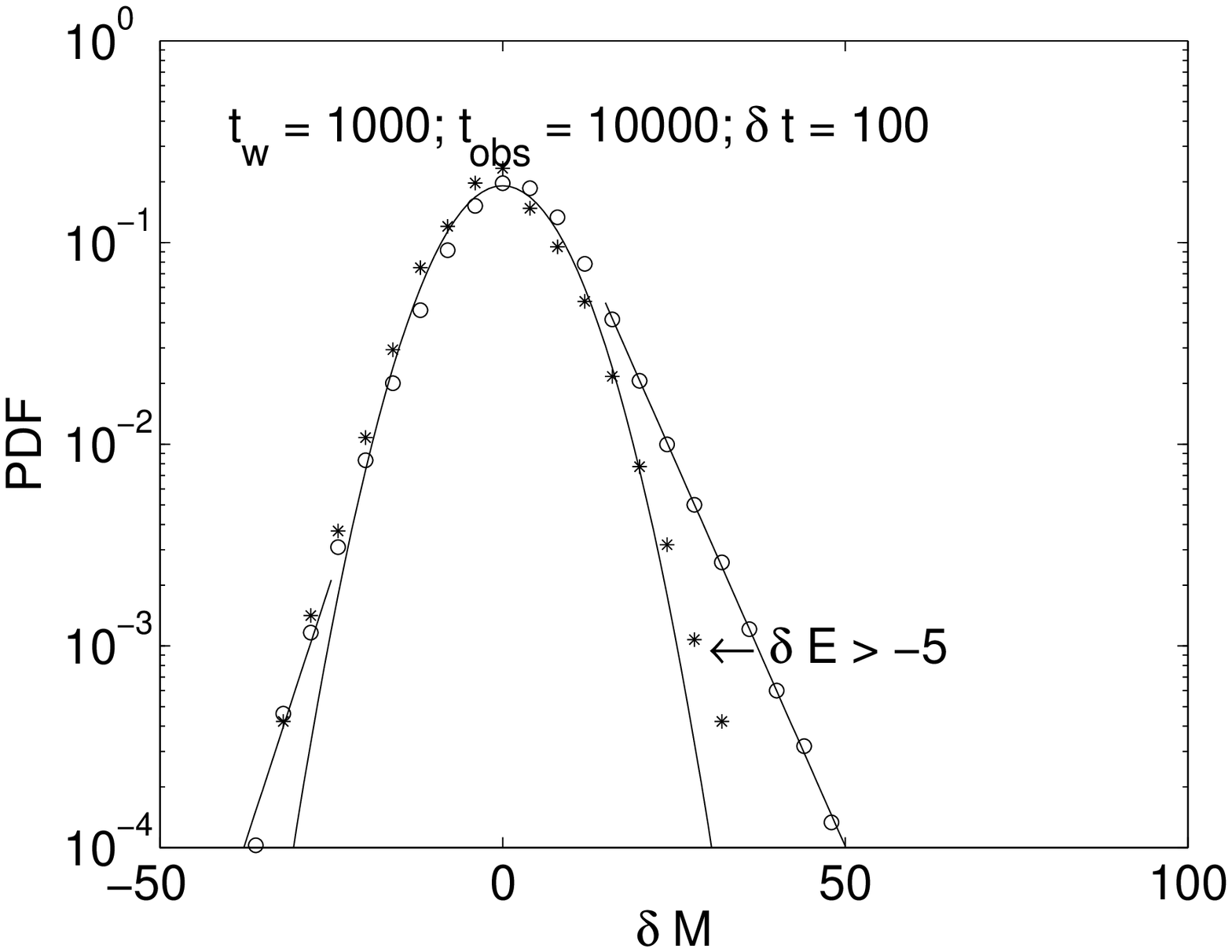,width=.47\textwidth}}  \quad 
}
\caption{(color on line) {\em (a)}:  The PDF of the spontaneous magnetic fluctuations
 has a Gaussian central part and two symmetric intermittent wings. Data are sampled
 in the interval $t_{\rm w}, t_{\rm w} + t_{\rm obs}$.
{\em (b)}: The same quantity  (outer graph) when  a field $H=0.3$ is turned on 
at $t_{\rm w} = 1000$. The left intermittent wing is clearly reduced, and the right intermittent
wing is amplified. The inner, almost Gaussian shaped, PDF  is the \emph{conditional} PDF obtained by
excluding   the magnetic fluctuations which happen in unison with the quakes. The lines are  fits to a zero centered Gaussian and to two exponentials, which  are  obtained using different subsets of the 
data.
 }
\label{mag_PDF}
\end{figure*}   
\section{Model and method} 
In spite of its simplicity  and, in particular, in spite of its lack of quenched disorder and
 the triviality  of its ground states, the p-spin model   has full fledged aging
behavior\cite{Lipowski00,Swift00,Sibani06b}.   
In the  model, $N$   Ising spins, $\sigma_i = \pm 1$ are  
placed  on a cubic lattice with periodic boundary conditions. They   interact through  
the plaquette Hamiltonian  
\begin{equation}
{\cal H} = -\sum_{{\cal P}_{ijkl}} \sigma_i\sigma_j \sigma_k \sigma_l +  H \eta(t_{\rm w} -t) \sum_i \sigma_i,
\label{hamilt}
\end{equation}  
where first sum  runs over all the elementary  plaquettes 
of the lattice, including for each the  product of the  four  spins 
  located at its corners. 
The second sum describes  the  coupling of the total magnetization  $\sum \sigma_i $ to  an external magnetic
field.  As expressed by the Heaviside step function $\eta(t_{\rm w} -t)$,
the field   jumps  at  $t=t_{\rm w}$  from zero to $H$.

  The simulations utilize the  
  Waiting Time Algorithm (WTM)~\cite{Dall01}, a rejectionless  algorithm
 endowed with an  `intrinsic'   time unit    
 approximately corresponding  to one Monte Carlo sweep.  
 By choosing a   high energy random configuration as initial state for low temperature isothermal
 runs, an   instantaneous  thermal quench is effectively achieved. 
All   simulations  temperatures are  within the model aging regime, $0.5<T<2.5$.
To improve the statistics, PDF data are collected over  $2000$ 
independent runs for each set of physical
parameters. Other statistical data are collected
 over $1000$ independent runs.
 
In the following, the symbol $t$ stands for the time elapsed from the initial quench 
(and from the beginning of the
 simulations).   The field is switched on at time  $t_{\rm w}$,
  and $t_{\rm obs} = t - t_{\rm w}$  is 
 the `observation' time, during which data are collected.  
The  external field jumps from zero to $H=0.3$ at  $t=t_{\rm w}$. 
  The thermal energy is denoted by $E$,   the magnetization 
 by $M$ and the fluctuations by $\delta E$ and $\delta M$ respectively. 
The latter are calculated as  finite time differences  over  time  intervals of length  $\delta t \ll t_{\rm obs}$. 
 The average magnetization per spin is  denoted by $\mu_{\rm ZFCM}$. 

\section{Results}
 That intermittent physical changes occurring 
in an   aging process  should be statistically subordinated to the quakes~\cite{Oliveira05,Sibani06a,Sibani07,Sibani06}
is a crucial element of   a record dynamics description. For completeness, we therefore 
include recent supporting evidence~\cite{Christiansen07} related to the present model.
 \begin{figure*}[t!]
$
\begin{array}{cc}
\includegraphics[width=0.45\linewidth,height= 0.4\linewidth]{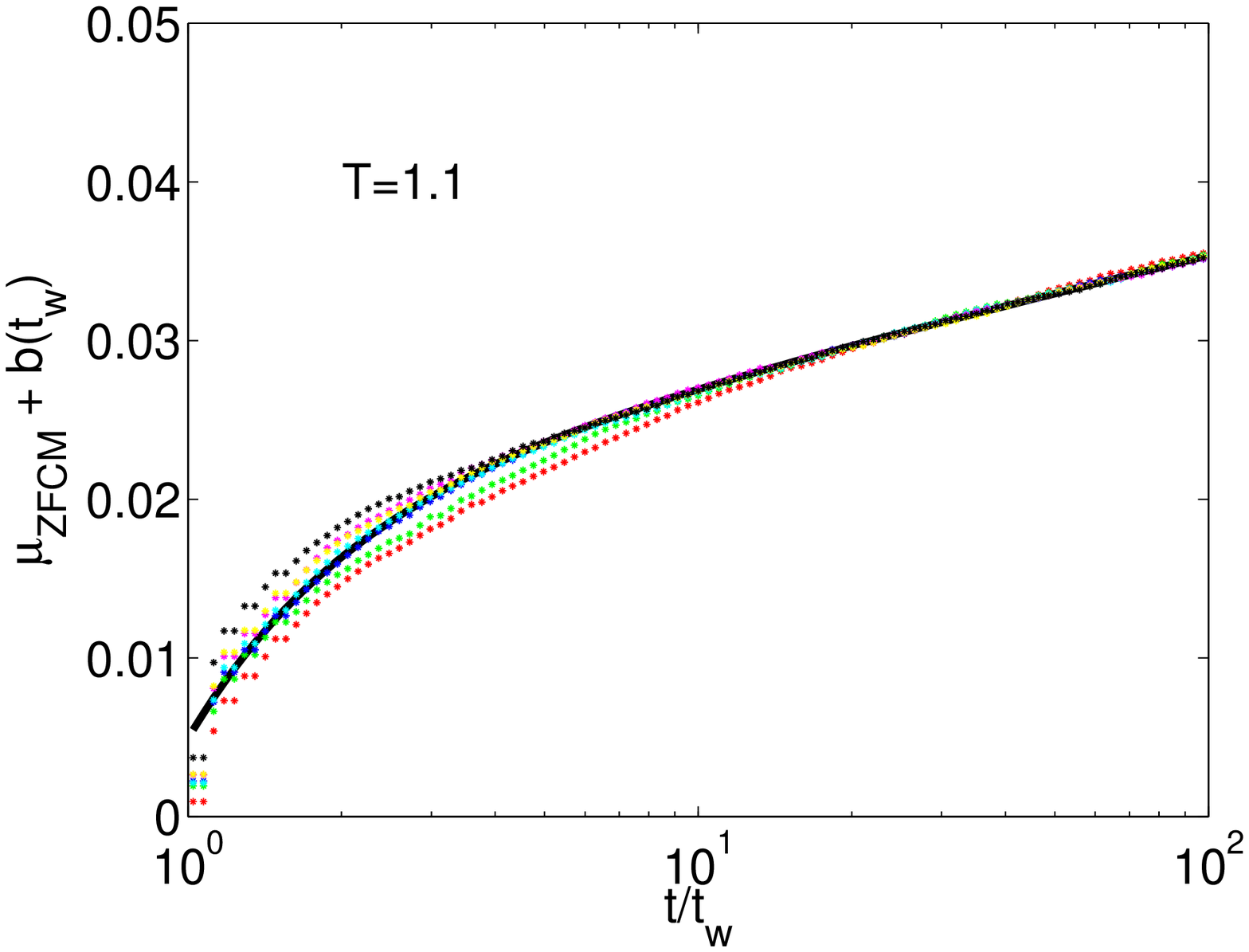}  &  
\includegraphics[width=0.45\linewidth,height= 0.4\linewidth]{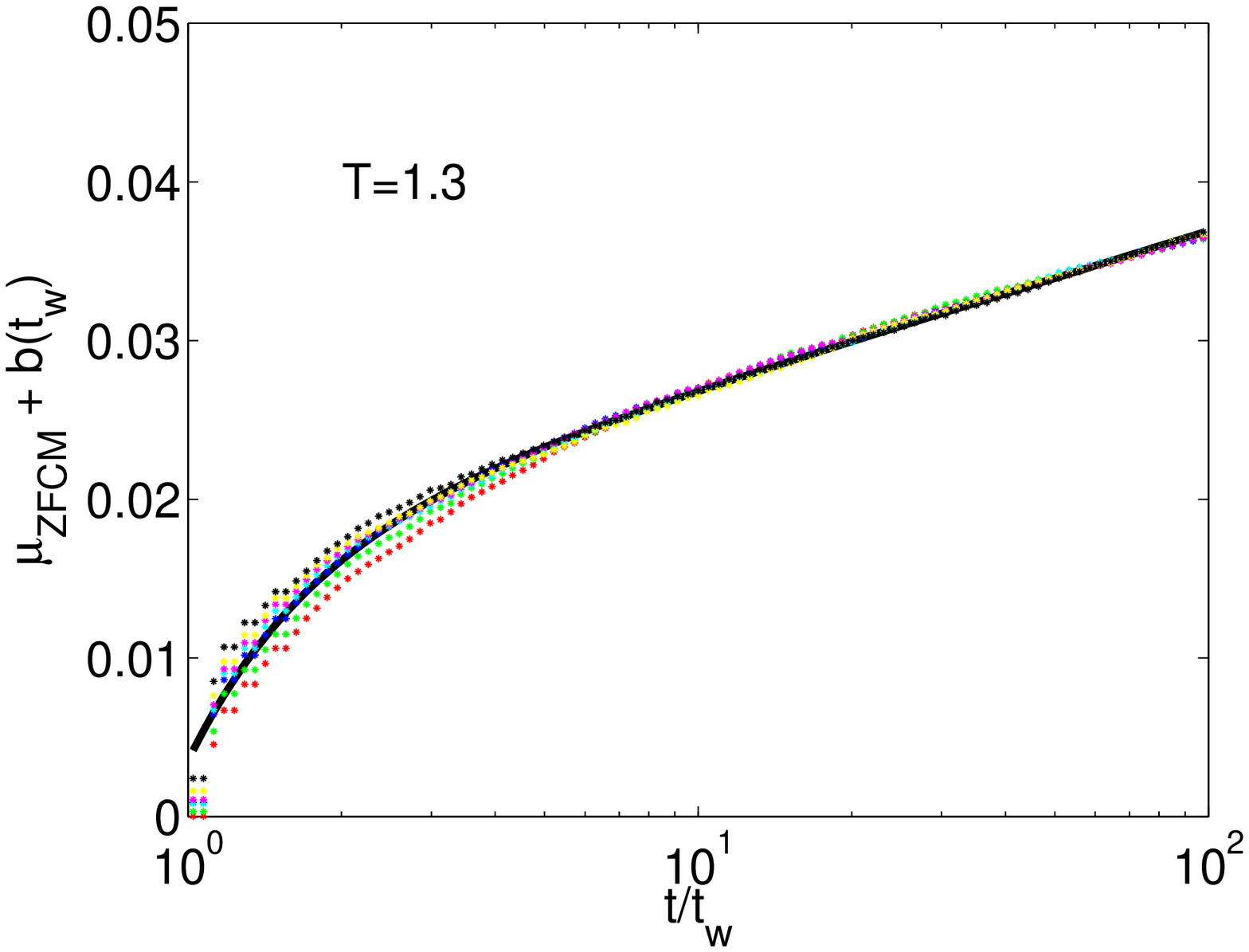} \\  
 \includegraphics[width=0.45\linewidth,height= 0.4\linewidth]{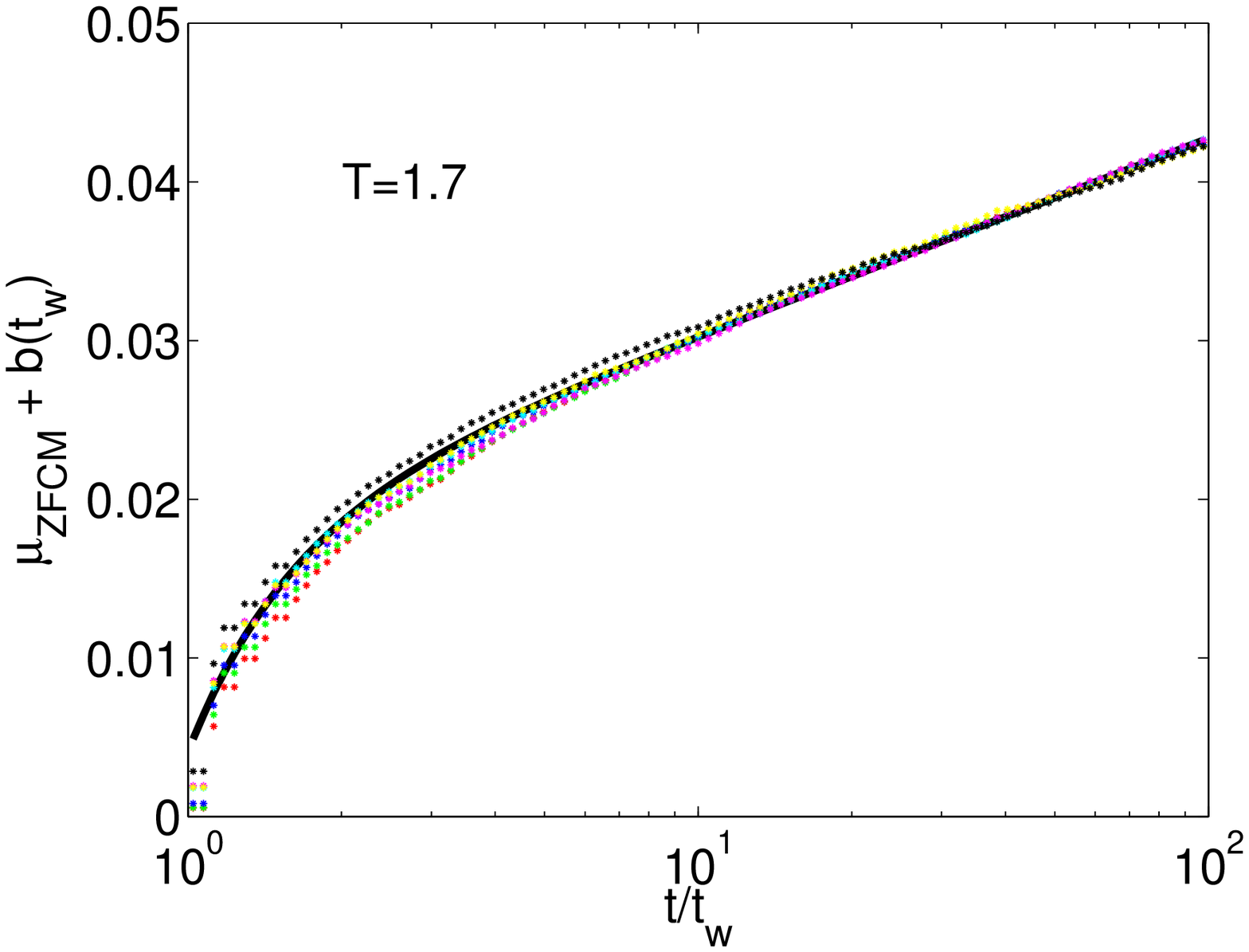}  &
 \includegraphics[width=0.45\linewidth,height= 0.4\linewidth]{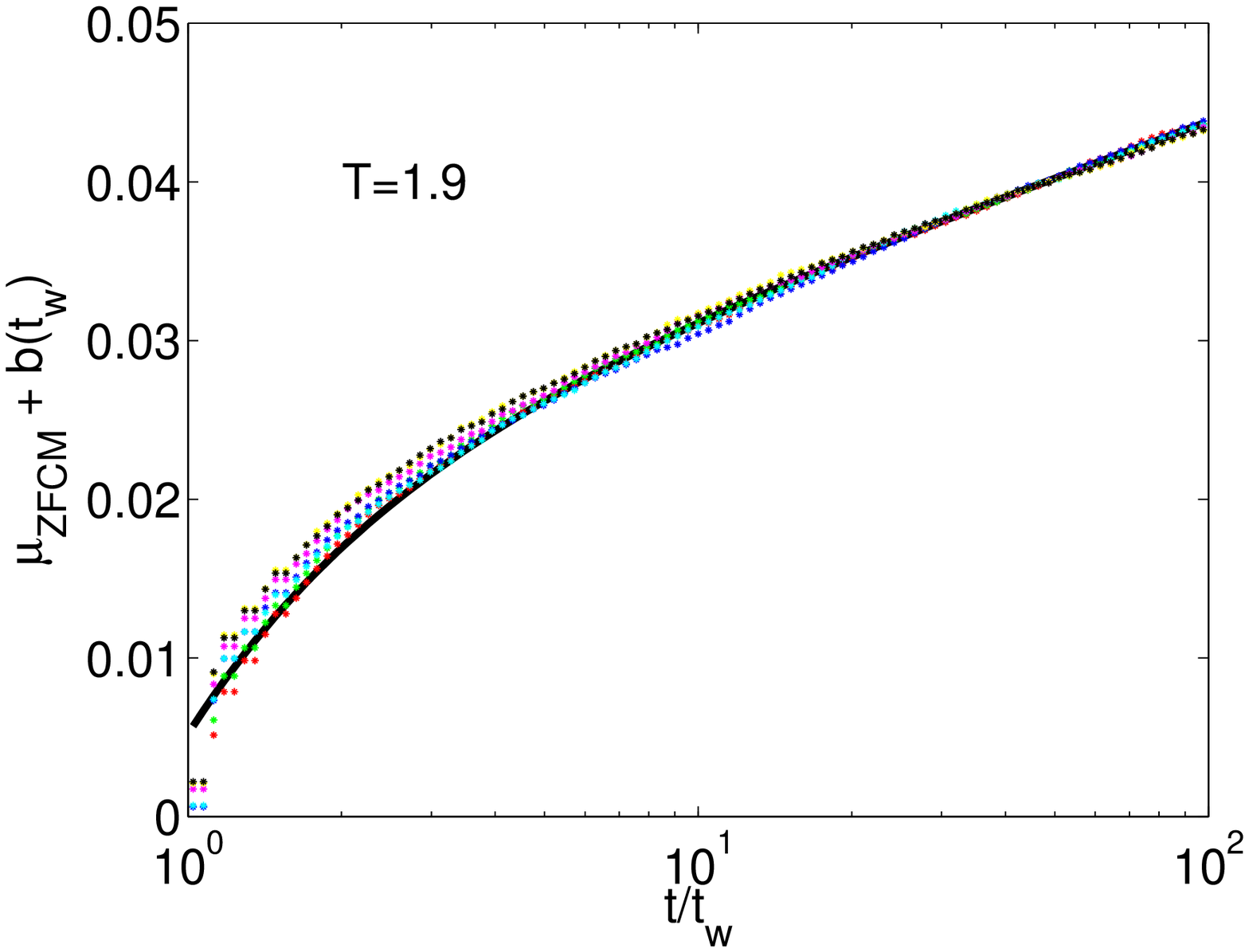}   
\end{array} 
$ 
\vspace{0.5cm} 
\caption{(Color on line)  For each  temperature  indicated,  the four panels depict  scaling plots 
of the average magnetization per spin, vertically adjusted  by the  constant  $b(t_{\rm w})$.  
   The values of $t$ range from $t=t_{\rm w}$ to  $400 t_{\rm w}$, with $t_{\rm w} = 50,100,200,500$ and $1000$, 
 (green, blue, magenta, black, and red circles respectively).
 The black line is   a fit to Eq.\eqref{alr}.  
}
\label{big_m}
\end{figure*}   

 The circles in Fig.~\eqref{mag_PDF} 
(see  Ref.~\cite{Christiansen07}) show, on a logarithmic vertical scale,     PDFs 
 of the  magnetic fluctuations $\delta M$  in the interval 
$[t_{\rm w},t_{\rm w}+t_{\rm obs}]$. The data in the left   panel  are  spontaneous magnetic fluctuations
(no field),  while those in the right panel are effected by  an external  magnetic field
  switched on at $t = t_{\rm w}=1000$. 
In both panels, intermittent wings   
extend a  zero-centered central Gaussian  spectrum. The  
 magnetic field  enhances  the positive wing,  reduces  the negative one, 
 and leaves    the Gaussian spectrum  unchanged. 
In the right panel,     a conditional PDF (stars) is also shown. 
The PDF  is   obtained by removing from the statistics all  magnetic fluctuations  occurring   
within the same  or the next   $\delta t$  as   
 energy fluctuations  of magnitude  $\delta E \le -5$.  As the    threshold chosen    
delimits the intermittent behavior of the heat flow PDF~\cite{Christiansen07},
the  filtering removes  the  magnetic fluctuations concurrent   with  the quakes. 
Since the resulting conditional PDF    is nearly Gaussian,   the  intermittent 
 magnetic fluctuations carrying the average magnetic response  are  
all  closely associated to the quakes. Furthermore, 
 the  energy, and,  in particular, the quakes, are    hardly affected
by the magnetic  field. Therefore,  the  quakes     dissipate the excess energy entrapped in the 
initial (zero field)  configuration~\cite{Sibani06b,Christiansen07}, and  
the field   probes but does not modify    the spontaneous intermittent dynamics.
 
As argued in ref.~\cite{Sibani03},  the number of quakes in the interval 
$(t,t')$ is  theoretically described by a Poisson distribution with average 
\begin{equation}
n_I(t',t)  = \alpha(N) \ln (t/t').
\label{average}
\end{equation} 
In the theory,  $\alpha(N)$ depends linearly on  the system size 
$N$ and  not at all on   temperature.
The  present model  
 has nearly  the same  behavior~\cite{Christiansen07}.
\subsection{Isothermal aging}
For physical processes  subordinated to statistically independent  quakes, 
 the number, $n$,  of quakes  falling in a  given  observation interval   
assumes the r\^{o}le of the  time variable in   a homogeneous Markov process.  
The  corresponding  moments   represent physical observables, and admit 
eigenvalue expansions. By  averaging such expansions over the Poisson distribution of $n$, 
 analytic formulas   are obtained  for 
 the moments~\cite{Sibani06,Sibani06a,Sibani07} which 
 depend  on the ratio $t/t_{\rm w}$ between the end-points,
$t_{\rm w}$ and $t$, of the observation interval. These  expressions 
 can usually  be   truncated after  a few leading terms. E.g. the average linear response of
the p-spin model is well approximated by  
\begin{equation}
\mu_{\rm ZFCM} = b_0 + a_m \ln(\frac{t}{t_{\rm w}}) + b_m (\frac{t}{t_{\rm w}})^{\lambda_m},
\label{alr}
\end{equation}
where $b_0$, $a_m$,  $b_m$ and $\lambda_m$  are  parameters. 
The exponent $\lambda_m$ is negative, whence, asymptotically, the 
behavior  becomes  logarithmic. 
Figure~\eqref{big_m} shows the average linear response versus $t/t_{\rm w}$
for the  temperatures indicated in the four plots. 
Each plot  displays data sets  (dots)  for $t_{\rm w} = 100,200,400,600,800,1000$ and $2000$.  The full line is
a fit to  Eq.~\eqref{alr}.    In order to  include      the stationary contribution to the magnetization,
which is ignored  in the theory, 
  the constant $b_0$ in Eq.\eqref{alr} is modified by an additive 
 $t_{\rm w}$ dependent term. The correction  increases with $t_{\rm w}$,
 but stays within a   few percent of the total. 
The black line is obtained by  fitting   to Eq.~\eqref{alr}. 
For the smallest values of the abscissa,   the quality of the data collapse, and consequently, of the fit,  is higher the higher the temperature. For larger values of the abscissa,  all fits are equally
satisfactory. In summary, the isothermal  aging behavior is  rather well described by 
Eq.~\eqref{alr}. 
\subsection{The effect of T-shifts} 
 The response function of spin-glasses~\cite{Granberg88,Komori00b} is systematically 
 affected by a small  temperature change, a so called $T$-shift,  applied during the aging process,
 usually together with a field switch at age $t_{\rm w}$.
  E.g. the  logarithmic derivative 
$S(t_{\rm w},t_{\rm obs})$ of the average magnetization as a function of 
$t_{\rm obs}$ peaks  at   $t_{\rm obs} = t_{\rm w}$ under isothermal 
conditions, and the peak  moves to   a 
lower, respectively higher, value  when  positive, 
respectively negative,  temperature shifts are applied.
As the same effect  can be achieved isothermally  at the final 
temperature  
using a shorter, respectively longer $t_{\rm w}$, an effective
age $t_{w,{\rm eff}}$  expectedly provides a   natural and general parameterization.
A study of the   intermittent heat  transfer of  the E-A spin glass
model~\cite{Sibani04a} under a $T$-shift, and the present study, confirm and  to some extent  
qualify   this conclusion.  

 \begin{figure*}[t] 
$
\begin{array}{cc}
\includegraphics[width=0.45\linewidth,height= 0.4\linewidth]{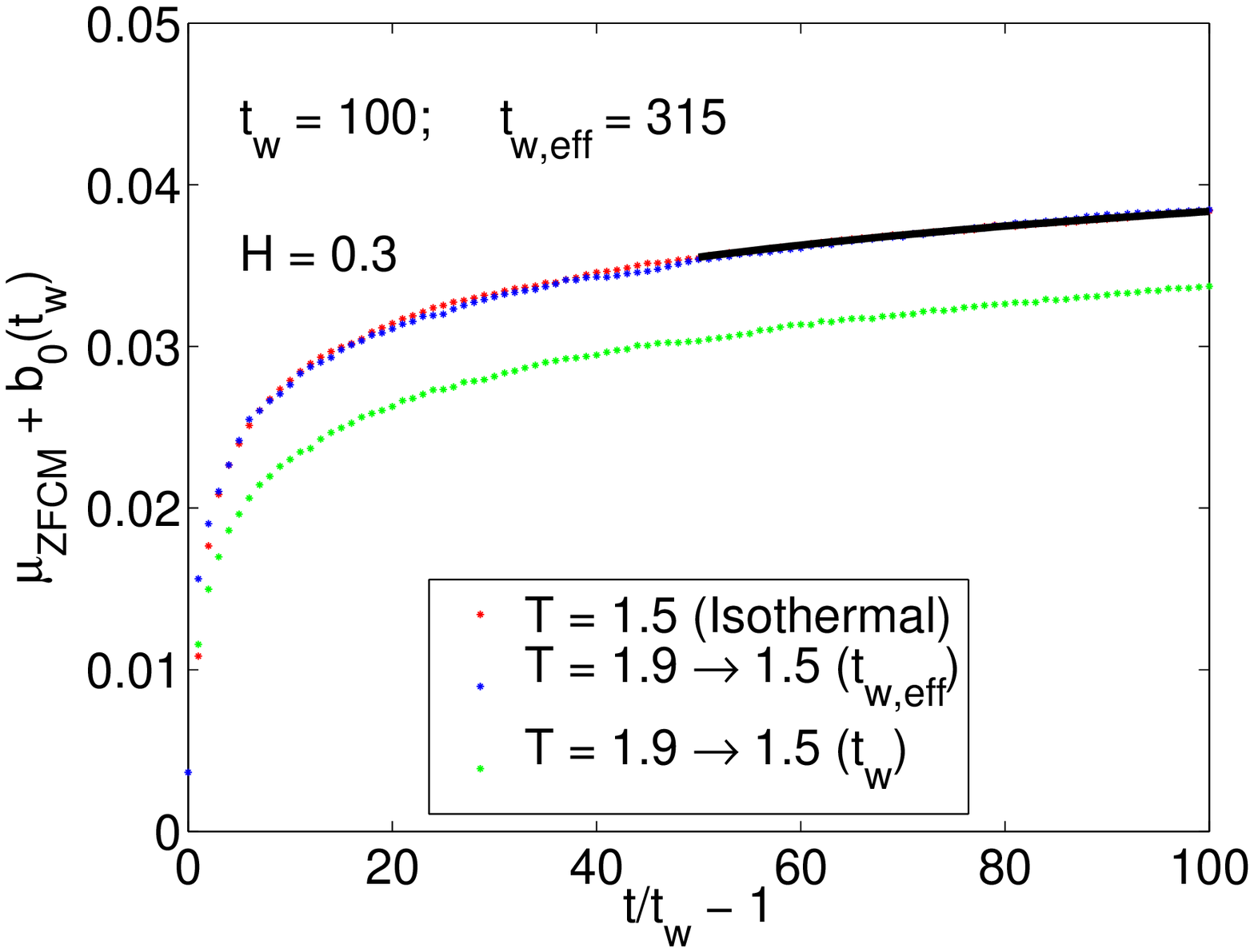}  &  
\includegraphics[width=0.45\linewidth,height= 0.4\linewidth]{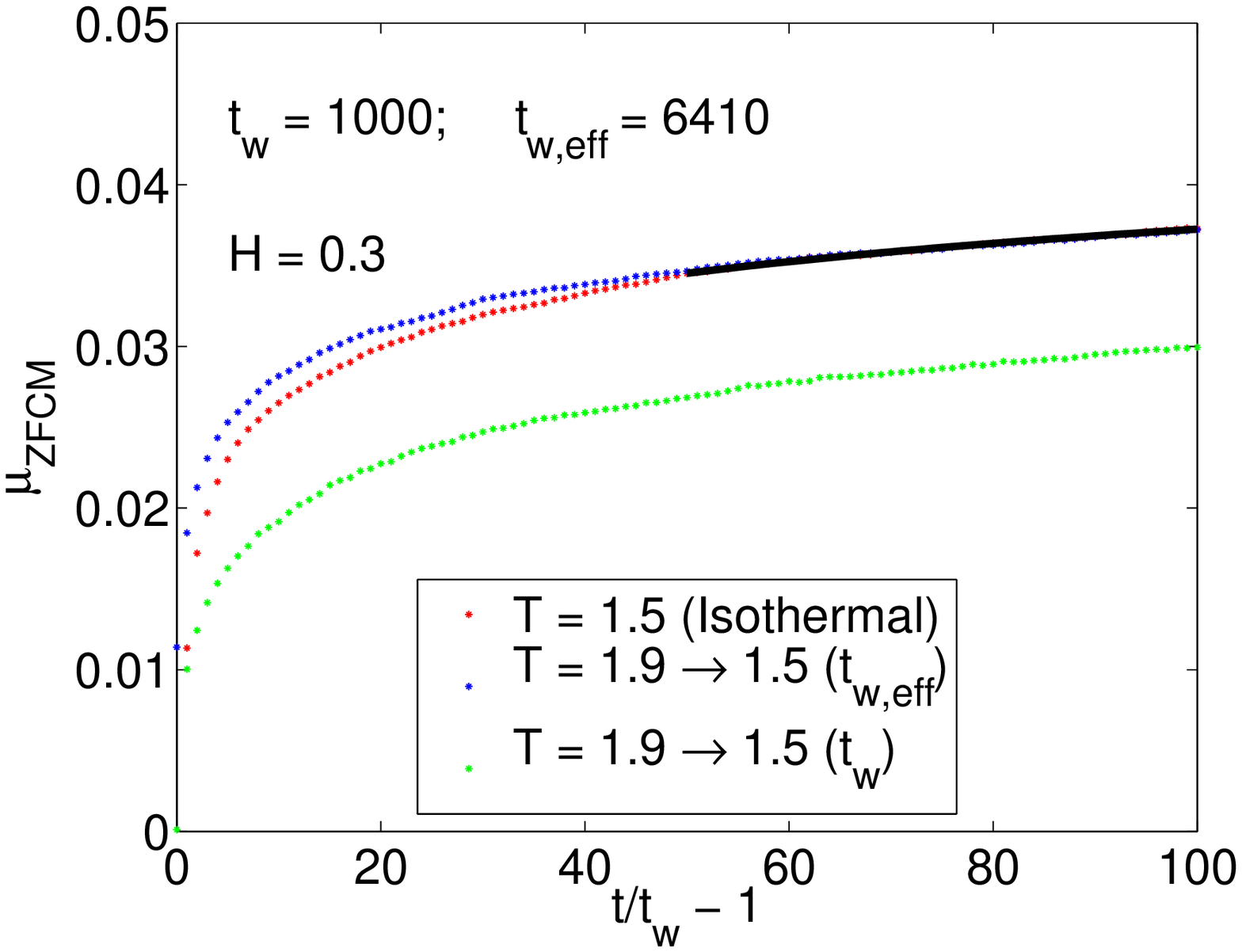} \\  
 \includegraphics[width=0.45\linewidth,height= 0.4\linewidth]{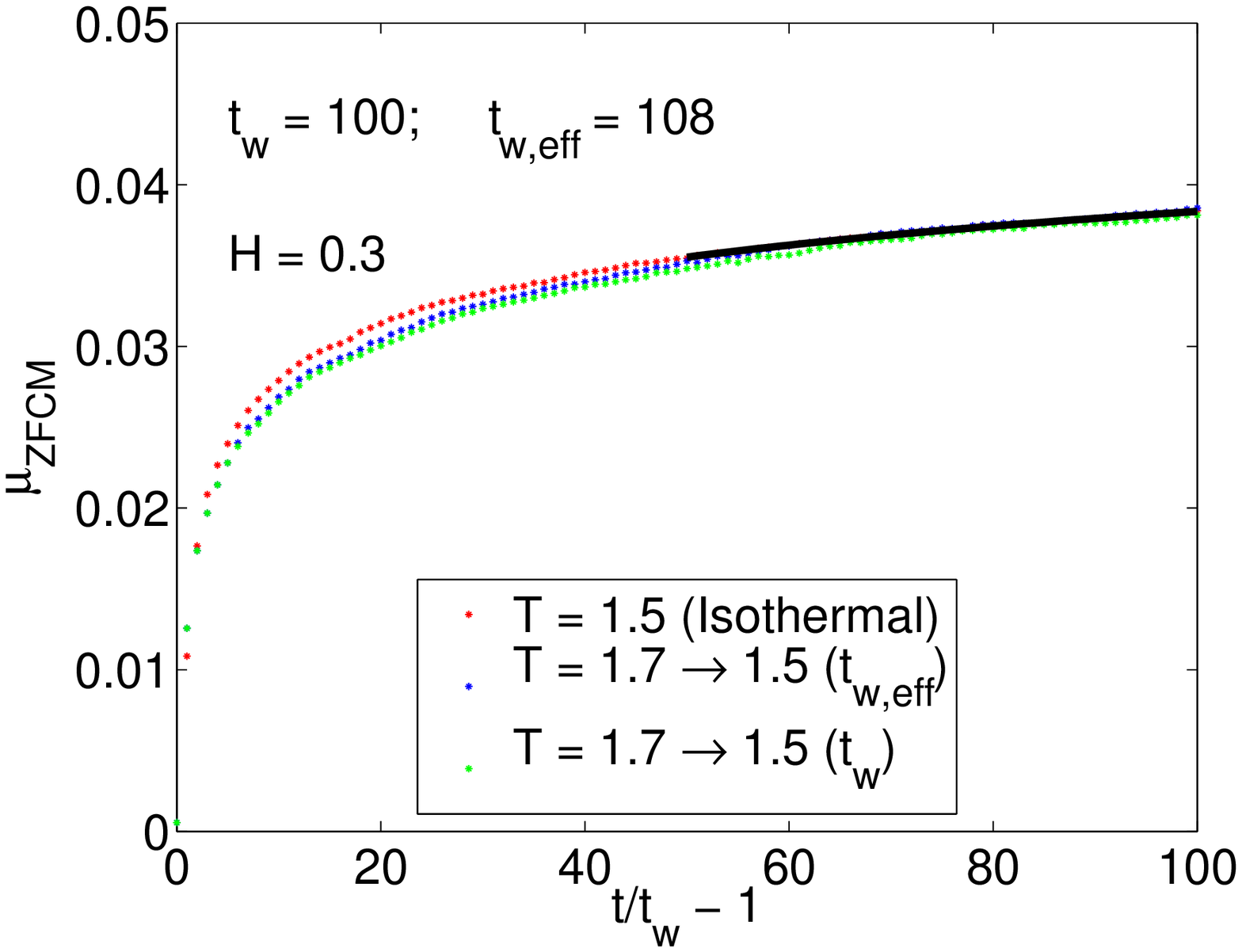}  &
 \includegraphics[width=0.45\linewidth,height= 0.4\linewidth]{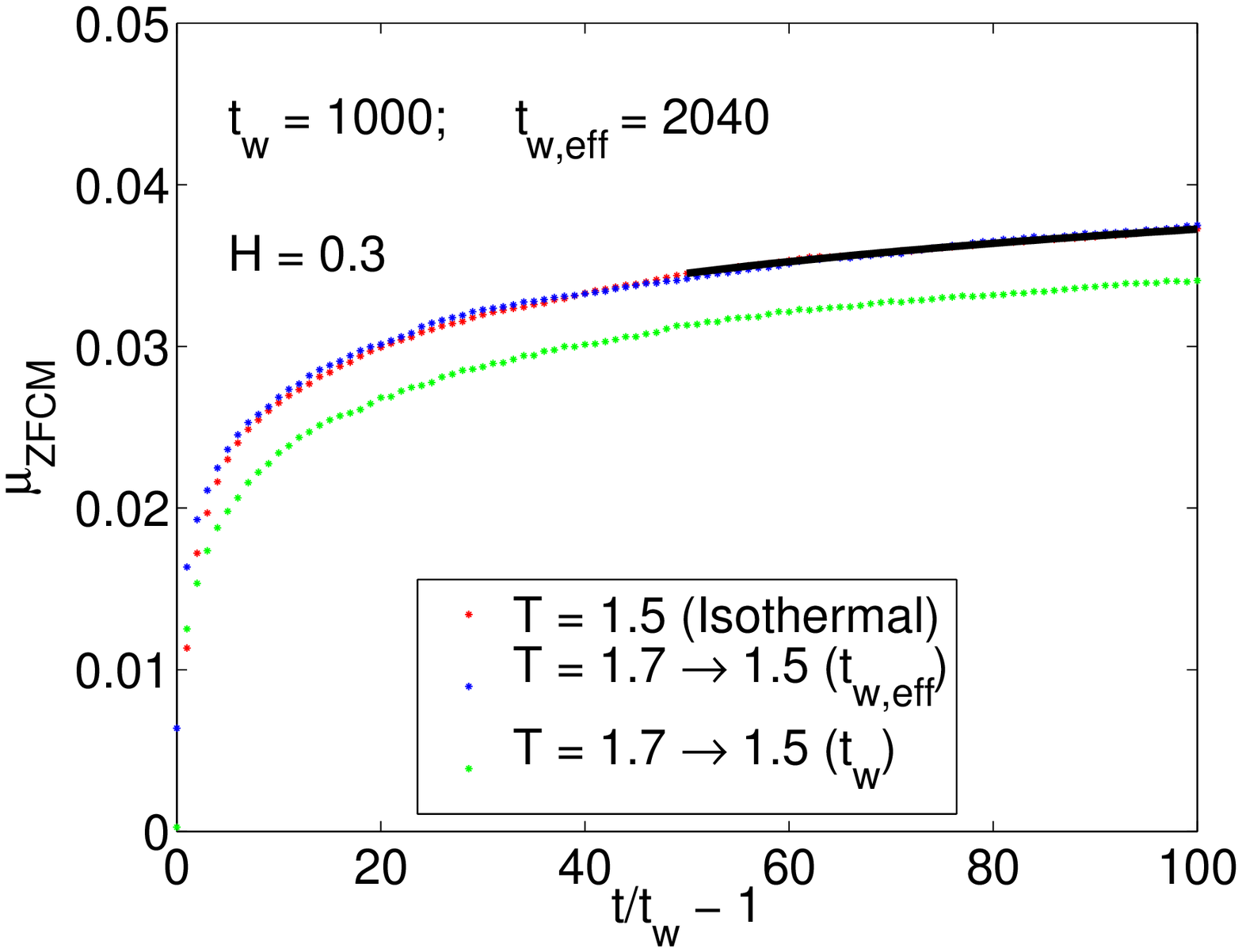}   
\end{array} 
$  
\caption{(Color on line)  For the  temperatures indicated, the four panels depict 
 scaling plots 
of the average magnetization per spin, shifted by the  constant  $b_0(t_{\rm w})$. 
In each panel,  three sets  of data are shown. The ordinate  is  the average linear response 
under different conditions. The abscissa is, a part from an additive constant,  the system age $t$,  scaled with either the true or the effective age.  The two nearly overlapping data sets are  for: \emph{i)} Isothermal response at the 
indicated  temperature,   versus the age scaled by  $t_{\rm w}$.  \emph{ii)}   $T$-shifted response, 
 plotted     versus the system age scaled by 
$t_{w,{\rm eff}}$ . 
The quality of the  collapse of these   two data sets  gauges  the
relevance  of the effective age parameterization.  \emph{iii)} Same data as \emph{ii)}, but using the actual  $t_{\rm w}$ value to scale the age. 
}
\label{big_shifted_m}
\end{figure*} 
  Below, we consider  
how both negative and positive   $T$-shifts  affect the 
average response function of the pla\-quet\-te model.  We furthermore discuss how    negative shifts affect  the 
spectrum of intermittent magnetization fluctuations.  
 \begin{figure*} 
 \vspace{2cm}
$
\begin{array}{cc}
\includegraphics[width=0.45\linewidth,height= 0.4\linewidth]{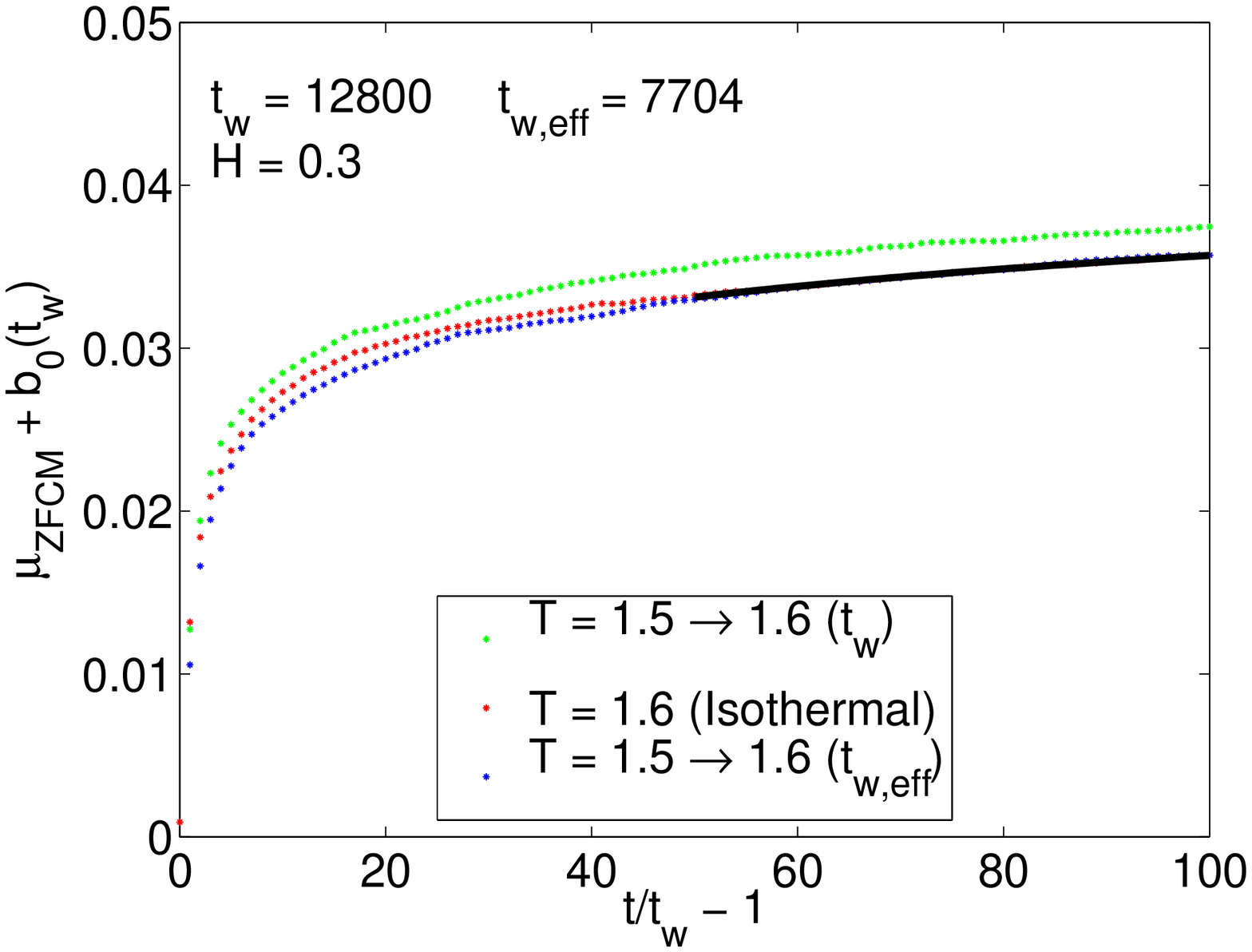}  &  
\includegraphics[width=0.45\linewidth,height= 0.4\linewidth]{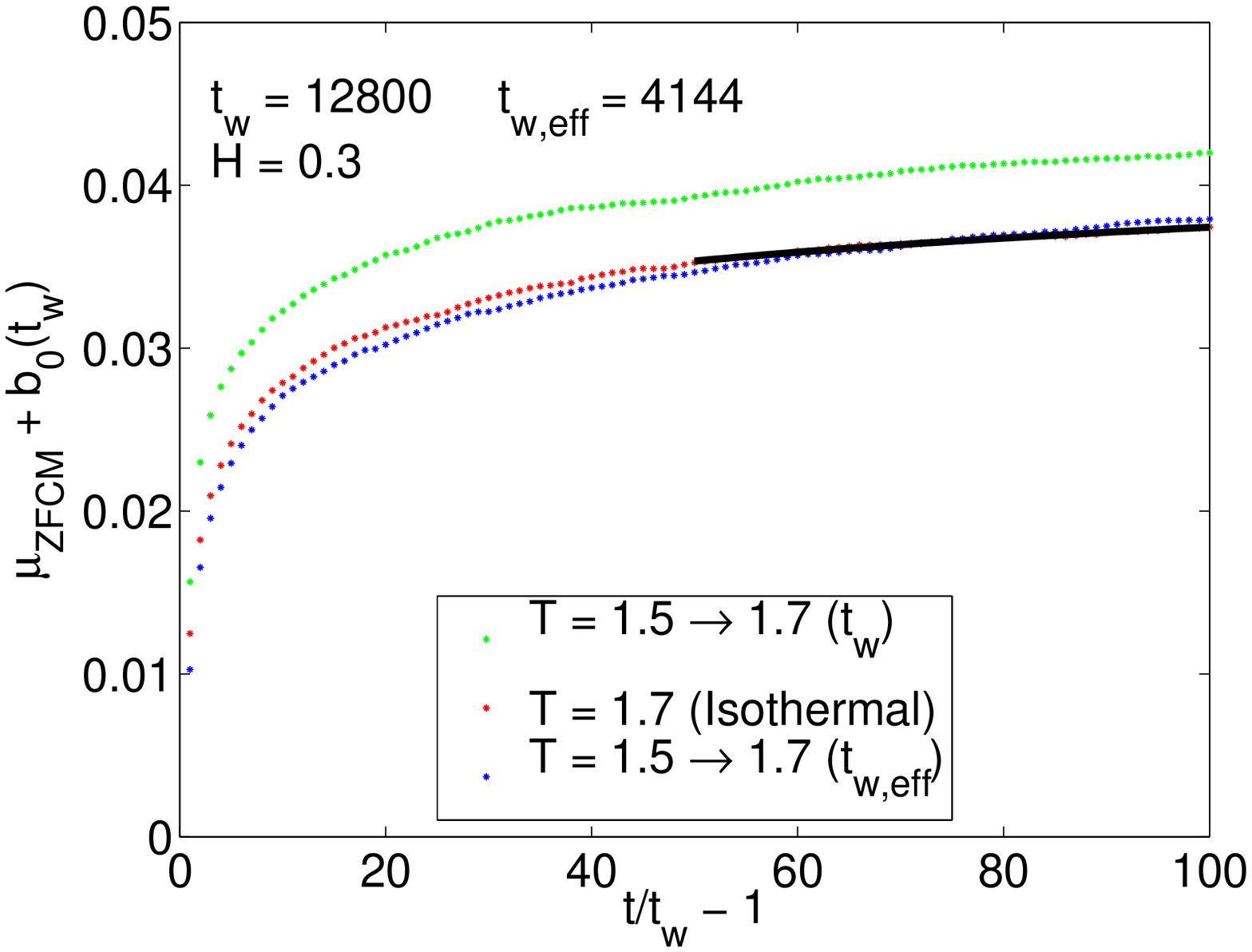} \\  
 \includegraphics[width=0.45\linewidth,height= 0.4\linewidth]{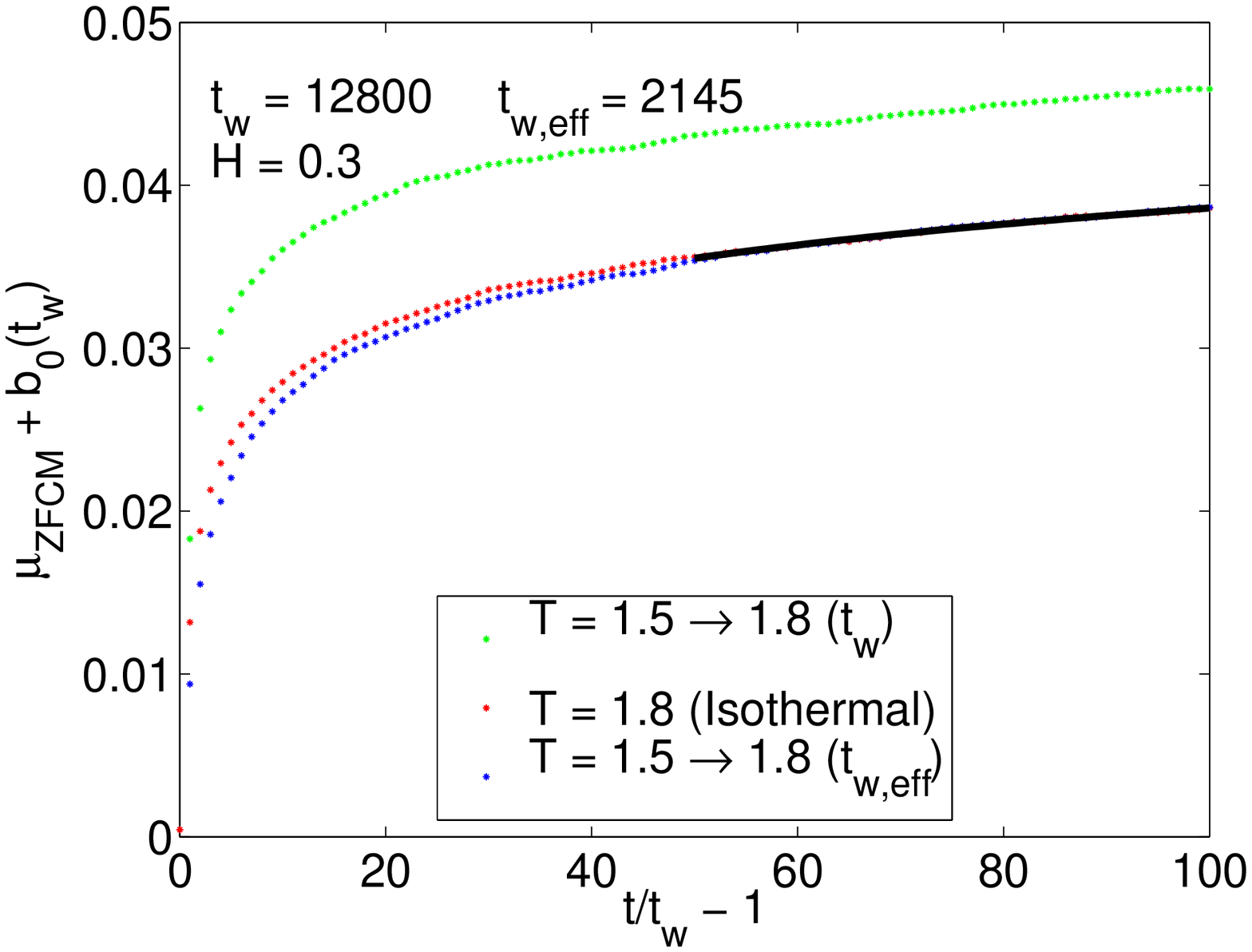}  &
 \includegraphics[width=0.45\linewidth,height= 0.4\linewidth]{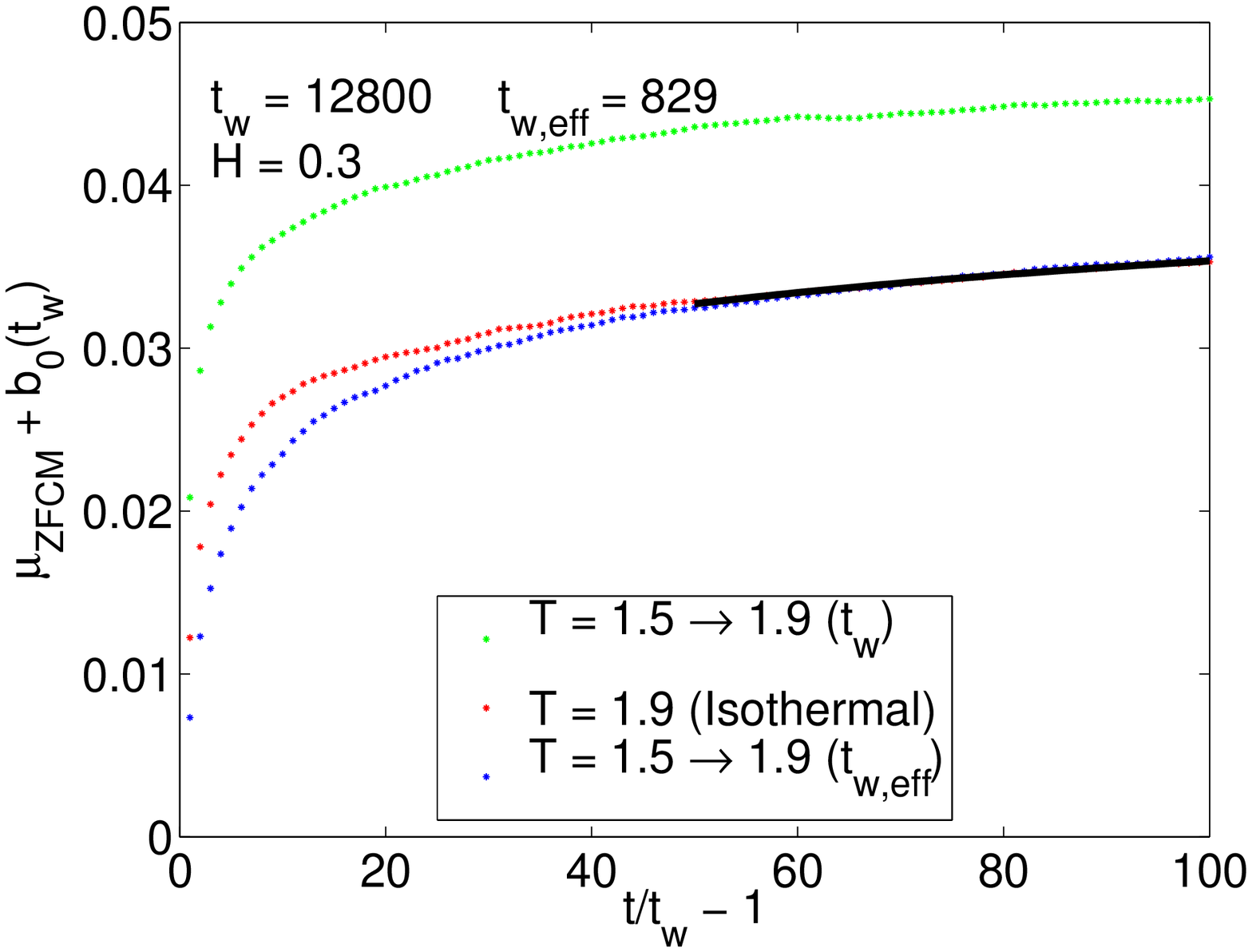}   
\end{array} 
$ 
\vspace{0.5cm} 
\caption{(Color on line) The four panels depict, for the  temperatures indicated, 
 scaling plots 
of the average magnetization per spin, vertically adjusted  by the  constant  $b_0(t_{\rm w})$. 
In each panel, three sets of data are shown. The ordinate  is  the average linear response 
under different conditions. The abscissa is, a part from an additive constant,  the system age $t$, 
scaled with either the true or the effective age.  
 The two  nearly overlapping data sets are  \emph{i)} isothermal response  at the
given temperature, versus the age scaled by  $t_{\rm w}$.  \emph{ii)}  $T$ shifted response. The
data are plotted  versus the age scaled by   
$t_{w,{\rm eff}}$. The quality of the  collapse of the  first two data sets  gauges  the
relevance  of the effective age parameterization. 
 \emph{iii)} Same data as \emph{ii)}, but using the actual $t_{\rm w}$ value to scale the age. 
}
\label{big_shifted2_m}
\end{figure*}   
Figure~\eqref{big_shifted_m} shows  the average magnetic
response,  vertically adjusted  by a small $t_{\rm w}$ dependent term as
previously discussed,   for a negative $T$-shift
 imposed concurrently  with the field switch at $t_{\rm w}$.
Each panel of Fig.~\eqref{big_shifted_m} contains three curves.
In the lowest curve   the  $T$-shifted  magnetization 
is plotted versus the scaling variable $t/t_{\rm w} -1$.
The two, nearly overlapping, and higher lying,  curves are 
\emph{i)} the   $T$-shifted data again,  now plotted versus $t/t_{w,{\rm eff}} -1$,
with   $t_{\rm w,  eff} $   chosen to  maximize  the 
 data collapse. \emph{ii)}   Isothermal magnetization at the final temperature,  plotted versus  $t/t_{\rm w} -1$. 
 Clearly, scaling by a suitable effective age  accounts rather well for the average linear response following a negative $T$-shift.
    \begin{figure}
$
\begin{array}{cc}
\includegraphics[width=0.9\linewidth,height= 0.8\linewidth]{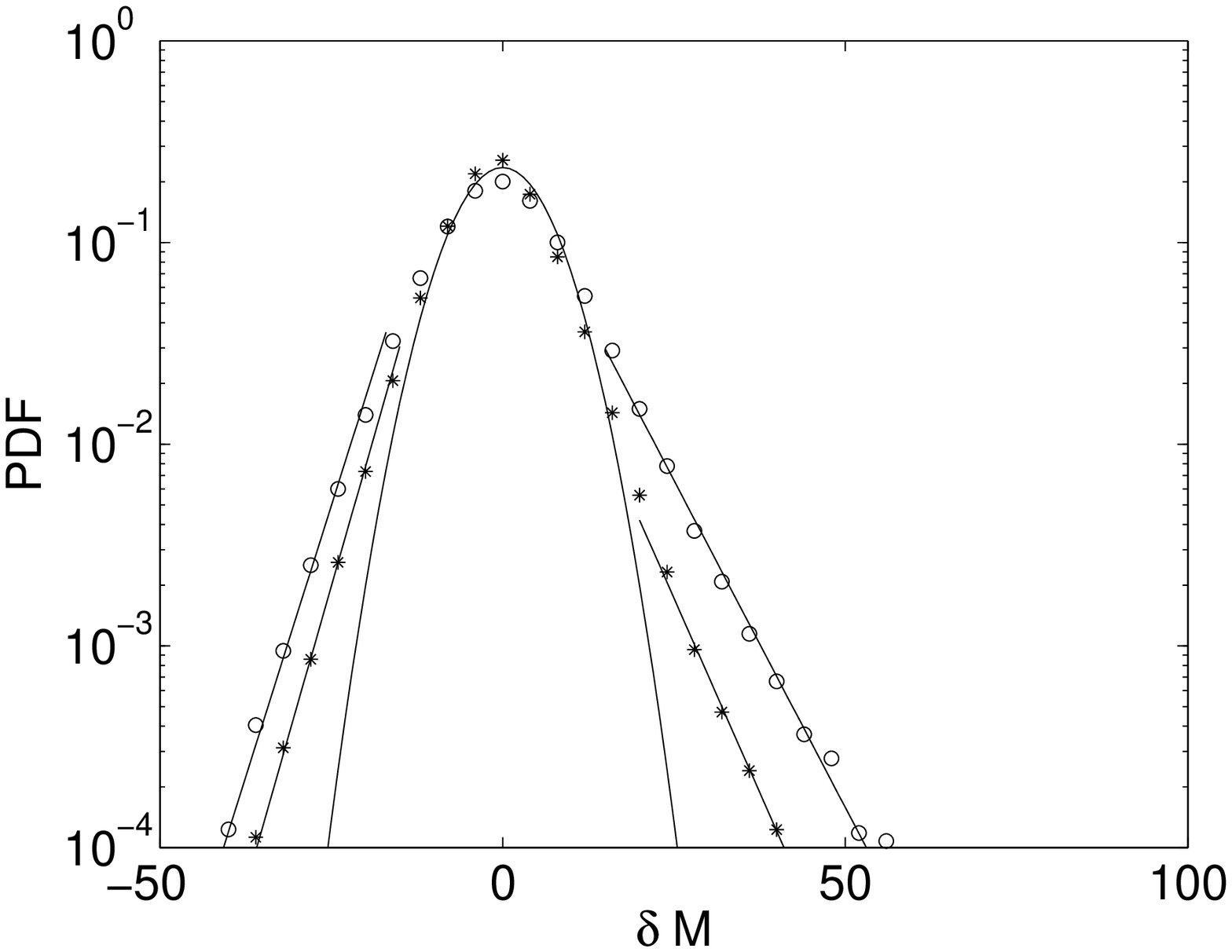}   
\end{array} 
$ 
\vspace{0.5cm} 
\caption{(Color on line) The PDF of magnetic fluctuations for isothermal (stars) 
and $T$-shifted (circles) protocols. The    data are collected in     intervals of type 
$[t,50 t]$. The initial ages are  $t_{\rm w, eff} =  6400$  for isothermal 
aging and   $t_{\rm w} = 1000$
for    $T$-shifted aging.  
The $\delta M$ values are   calculated over 
intervals  of size $\delta t=500$. A full collapse would imply the  full equivalence
of the   $T$-shifted response  and the isothermal response of a system of the 
 correct effective age.  The lines are  fits to a zero centered Gaussian and to two exponentials, which  are  obtained using different subsets of the 
data.
}
\label{SHIFTED_PDF}
\end{figure}   
 Figure~\eqref{big_shifted2_m} is similar to Fig.~\eqref{big_shifted_m},   
 except that   the applied $T$ shifts are   now  positive.
In each  panel, the highest lying curve shows  the  $T$-shifted 
magnetization versus $t/t_{\rm w} -1$.   
The two lower  lying curves are 
\emph{i)} the  $T$ shifted magnetization data,  now plotted versus $t/t_{\rm w,  eff}  -1$,
and, \emph{ii)}  the magnetization  curve  obtained for  isothermal  aging at the final temperature,  
plotted versus  $t/t_{\rm w} -1$. Note that the transient effects produced by a positive shift
are  not  fully   described  by  an  effective age. Consequently,  the    
  `best'  value of  $t_{\rm w,  eff} $  is obtained by a fit 
which only maximizes  data overlap for large $t/t_{\rm w}$ values. 
 
 Figure~\eqref{big_shifted2_m} shows that  positive $T$ shifts  lead, as expected,  to  effective ages  smaller than the 
  actual ages.
 The best achievable data collapse    is not as good 
 as  for negative shifts, especially for small  values
of $t/t_{\rm w}$.  Since   a  positive
 $T$-shift destroys, totally or in part, the current configuration, it  apparently  induces  additional 
 dynamical effects not well  represented by an   effective age. 
Qualitatively, the  numerical  results 
 for positive shifts    concur  with a crucial  hypothesis of  record dynamics, namely that the 
 attractors  successively visited are all marginally stable,  i.e. they are destabilized  by a record sized
 thermal fluctuation. Increasing the temperature leads to larger thermal fluctuations,  and hence
 to the  destruction of the configuration reached at $t_{\rm w}$.  
Negative    shifts  have  more subtle consequences,  which  yield to an  analytical description.
 We first consider how negative shifts   affect   the   spectrum of the magnetization fluctuations.
  Secondly, the algebraic relationship between true and effective age is explained theoretically and checked 
  against simulation data.
 
A system undergoing a   negative  $T$-shift  $T=1.9 \rightarrow T=1.5$ at $t_{\rm w} = 1000$ has,
according to  Fig.~\eqref{big_shifted_m}, second
panel,  an effective age    $t_{\rm w,eff} \approx 6400$ .   
If its  dynamics were fully equivalent to the isothermal dynamics of 
a system of age $t_{\rm w,eff} \approx 6400$,  the 
 PDFs of   magnetization differences $\delta M$ 
calculated    over intervals of the same length $\delta t$ would overlap in the two cases.   
In    Fig.~\eqref{SHIFTED_PDF}, the   PDF of isothermal fluctuations for a system with $t_{\rm w} = 6400$
 (stars), are compared to 
 the corresponding PDF  of a $T$-shifted system of age $t_{\rm w} = 1000$ (circles).
 Both data sets  are plotted on a  logarithmic vertical  scale.  The Gaussian parts overlap, 
but the  intermittent tails  do not. They are however parallel, meaning that   the weight of the intermittent 
fluctuations relative to the weight of the Gaussian fluctuations  is  higher 
in the $T$-shifted than in the isothermal case, even though  the normalized   
size distribution of  the intermittent fluctuations is the same in both  cases.
 Since  the Gaussian fluctuations  have zero average,  the average response is described via $t_{\rm w,eff}$ as previously discussed. 
 \begin{figure*}
$
\begin{array}{cc}
\includegraphics[width=0.45\linewidth,height= 0.4\linewidth]{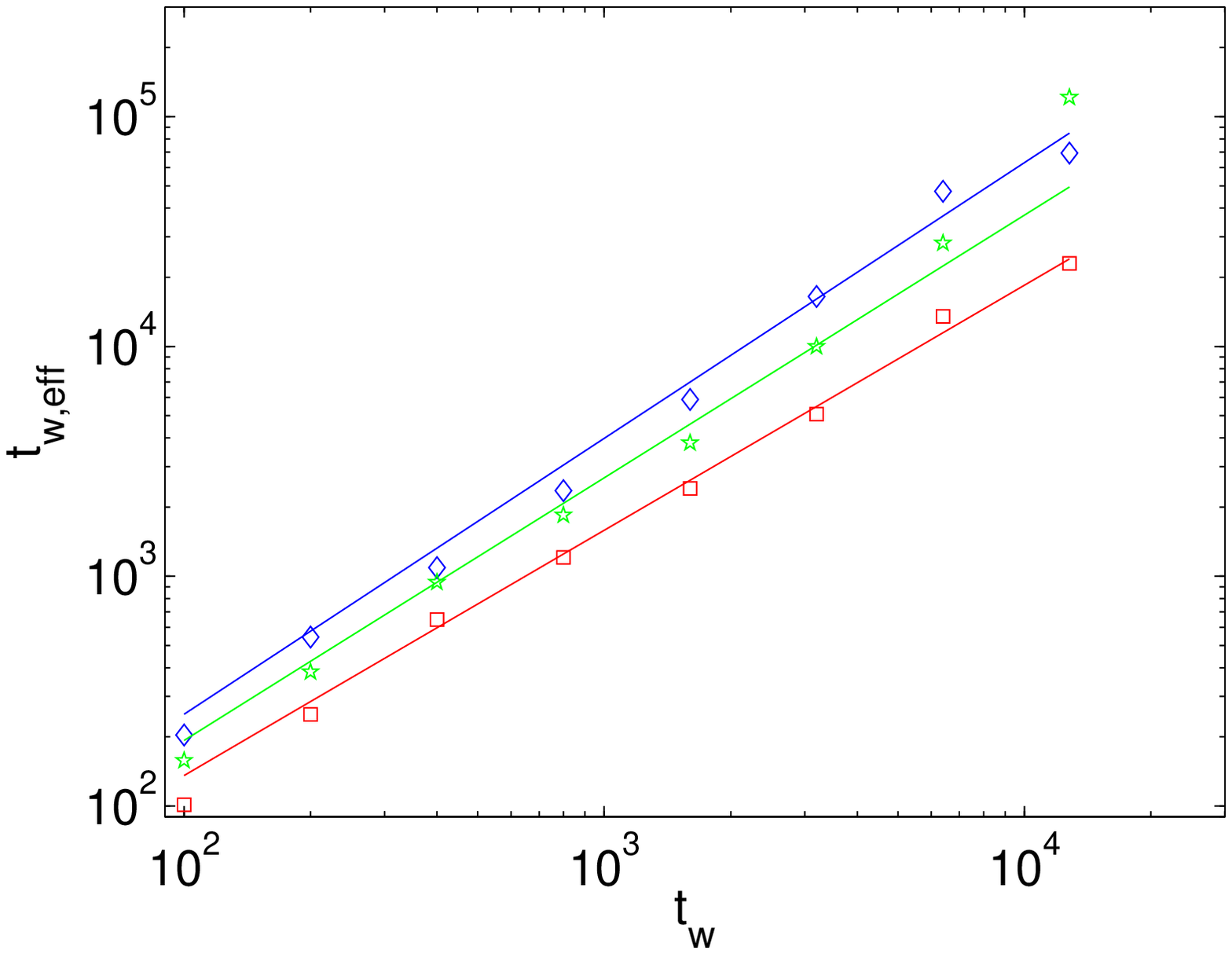}  &  
\includegraphics[width=0.45\linewidth,height= 0.4\linewidth]{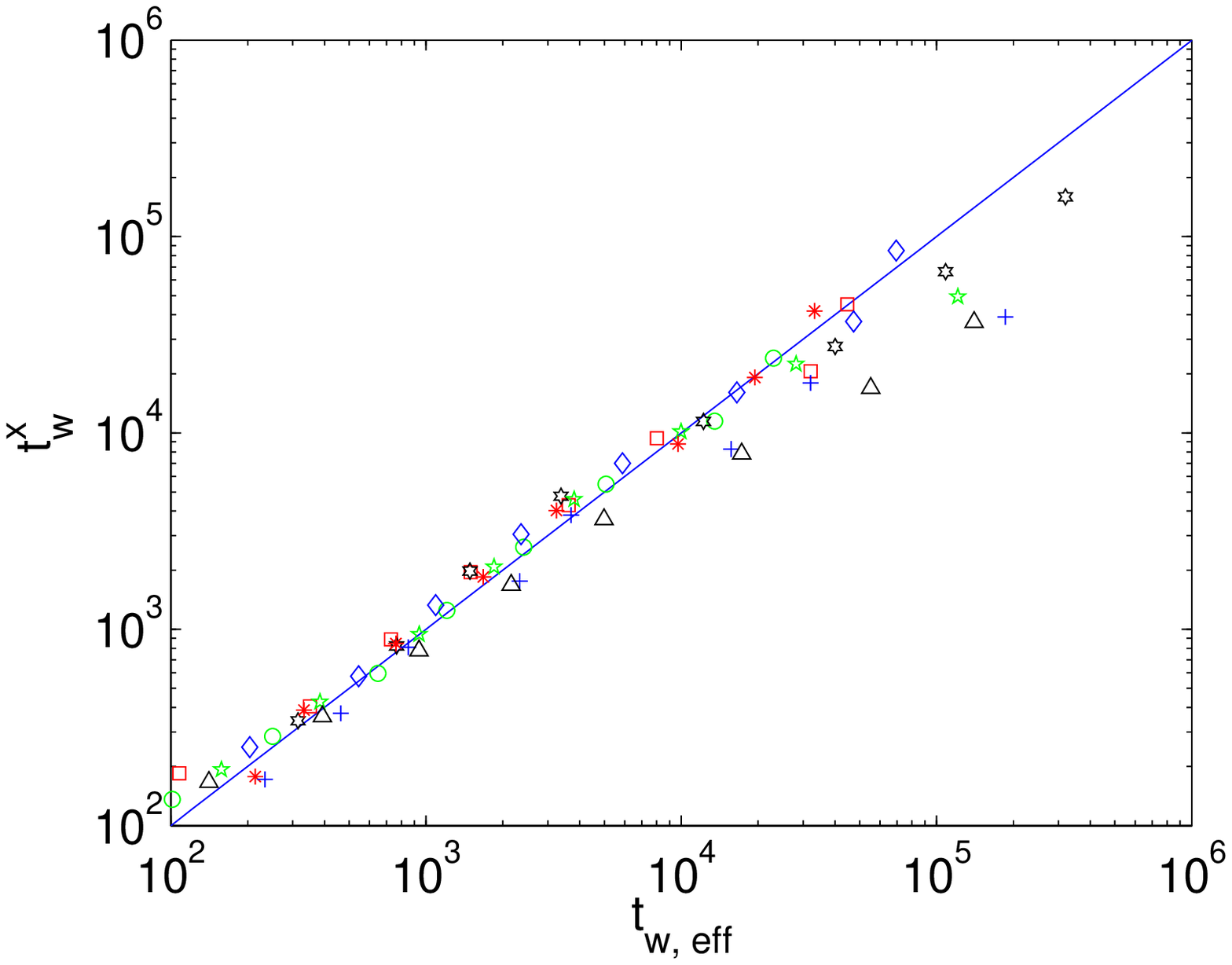}  
\end{array} 
$ 
\vspace{0.5cm} 
\caption{(Color on line) The left panel shows the algebraic relationship between the 
effective and the actual age for the three different   $T$-shifts 
 $1.6 \rightarrow 1.5$ (red squares),
$1.8 \rightarrow 1.5$ (green polygons) and $1.6 \rightarrow 1.4$ (blue diamonds). 
Other cases analyzed are not shown for graphical reasons
The full lines  correspond to  the analytical prediction  $t_{\rm w, eff} = t_{\rm w}^x$, where
$x$ is determined by a least square fit. 
In the  right panel,    $t_{\rm w}^x$ is plotted versus $t_{\rm w, eff}$  using different 
symbols  and colors for all available $x$ values. 
Deviations from the   line   gauge the discrepancy between  predicted and observed behavior.
}
\label{tw_vs_tweff}
\end{figure*}  

In  a record dynamics scenario, the  marginally stable attractor reached at $t_{\rm w}$
is associated to a dynamical barrier of magnitude   $b(t_{\rm w})$, which  matches 
 the largest among the thermal  fluctuations 
experienced by the system during the aging process. 
  Thermal  fluctuations    are  described by the  equilibrium Boltzmann  distribution
 $P(E)$ characterizing  a single thermalized domain. In our  case, the local density 
 of states can be treated as a constant, and   $P(E) \propto \exp(-E/T)$. The typical  value of the 
 largest   among  $t_{\rm w}$ fluctuations at temperature
 $T$  then scales  as $b(t_{\rm w}) \propto T \log (t_{\rm w})$~\cite{Sibani04a}.  Note that the 
 Arrhenius escape  time 
 corresponding to   $b(t_{\rm w}) $     equals  $t_{\rm w}$.    
 \footnote{ In the Edward-Andersen spin-glass model, the  local equilibrium distribution  and  the  exponent $x$ are
  affected  by   an exponentially growing local density of states ${\cal D}(E) \propto  \exp(E/  \epsilon)$. 
 In the  present case the local density of state has no important energy dependence.}.   
  After a  negative $T$-shift from $T_i$ to $T_f < T_i$,  the Arrhenius time corresponding to the  barrier 
  $b(t_{\rm w})$ becomes $t_{\rm w, eff} = \exp(b(t_{\rm w})/T_f) = t_{\rm w}^{T_i/T_f}$. 
  We thus  arrive at the conclusion that 
 \begin{equation}
 t_{\rm w, eff} = t_{\rm w}^{x(T_i,T_f)}; \quad x(T_i,T_f) = \frac{T_i}{T_f}.
 \label{th_eff_age}
 \end{equation}
 Note that the temperature dependence of the exponent  $x$ given by the above formula has no adjustable parameters.
In Fig.~\eqref{tw_vs_tweff}, left panel,   the empirically determined effective age is plotted 
versus  the actual age on a log-log scale, for  $T$-shifts    $ 1.6 \rightarrow 1.5$ (red squares)
 $ 1.8 \rightarrow 1.5$ (green polygons) and  $1.6 \rightarrow1.4$ (blue diamonds). The corresponding
 lines are obtained by linear regression, and their slopes provide a numerical estimate of the exponent $x$.  
Several other $T$-shifts  are also considered, but the corresponding data are  not shown  for graphical reasons. 
The values of   $t_{\rm w}^x$  corresponding to all available  $x$ values   are   plotted 
 versus the empirical effective age in the right panel of Fig.~\ref{tw_vs_tweff}.
 The color and symbol coding is as follows: 
$1.6 \rightarrow  1.5 $ (green circles), $1.7 \rightarrow  1.5 $ (red squares), $1.8  \rightarrow  1.5 $ (blue diamonds)
 $1.9 \rightarrow  1.5 $ (black hexagons),  $1.6 \rightarrow  1.4 $ (green polygons),
$1.8  \rightarrow  1.6 $ (red stars), $1.8 \rightarrow  1.6 $ (blue plusses) 
and $2.0 \rightarrow 1.8$,  (black upper triangles). The full line is the 
theoretical prediction given in Eq.~\eqref{th_eff_age}.  
Thus, the distance  between the data points
and  the line represents the mismatch  between  data and theory.  
Except for shifts involving a relatively high initial temperature, especially for high values of the effective age,
the agreement is good. 
\section{Discussion}  
Summarizing the results of Refs.~\cite{Oliveira05,Sibani06,Sibani06a,Sibani07},
using the number of quakes $n$  as an effective `time'  variable   produces
a     translationally invariant  re-parameterized aging description. 
This  description  relies on   the statistical independence of the quakes and 
on the statistical subordination of physical changes to the quakes.
It does not  rely  on the amount of   change  induces by the events.
 The  (stochastic) changes  may 
 depend  on the system type, on the observable considered,   
 and even on $n$ itself. They can either be  dealt with   
with   further modeling assumptions~\cite{Oliveira05,Sibani06}, or 
 using the generic property that
 physical averages   admit  eigenvalue expansions, when treated as a function of $n$~\cite{Sibani06a}. Averaging  such expansions over  the Poisson distribution 
of $n$ turns   each  exponential term   e.g. $a e^{c n}$, 
into  a fractional power   $ (t/t_{\rm w})^{\lambda_c}$,  where the  exponent $\lambda_c$  is
related to the eigenvalue $c$ of the time re-parameterized problem by 
\begin{equation} 
\lambda_c(T) = \alpha(N)(e^{c(N,T)} -1).
\label{exponent} 
\end{equation}
The  dependence of the exponent on the system size $N$ and temperature $T$  is now  explicitely
introduced. Note that  
the symbol $\lambda_c$  here  denotes   a generic exponent, 
and is not to be confused with the (nearly)  homonymous  dynamical exponent
$\lambda$    widely used in the literature. 
Considering that   $\alpha(N) \propto N$, several exponents  may  exist
which   diverge with  $N$. These exponent are 
dynamically irrelevant as the corresponding modes
quickly decay to zero. Some exponents may remain bounded in the thermodynamical
limit, namely if the  eigenvalue of the time-re-parameterized problem, $c(N)$
vanishes as $1/N$ or faster, as $N \rightarrow \infty$. Eigenvalues approaching 
zero faster than $1/N$  correspond to frozen modes, and are also dynamically 
irrelevant.  
Observable power-laws are thus connected to   relaxation `time' scales   of the 
re-parameterized  dynamics   which diverge 
linearly with   system size, see e.g.  Ref.~\cite{Sibani06} for an explicit demonstration
of such  linear divergence.   
The  corresponding  exponents  have  the 
 form   $\lambda_c(T) = \alpha(N) c(N,T)$,   which is    asymptotically independent of $N$ for large $N$.
 
Far away from equilibrium, and hence far  from saturation,  the smallest 
  relevant  exponent   can be small enough to justify  the further  expansion 
 $(t/t_{\rm w})^{\lambda_c} \propto 1 + \lambda_c \ln (t/t_{\rm w})$ for a wide range of $t/t_{\rm w}$. 
 If a logarithmic time dependence exists, it will be dominant 
 for large values of $t/t_{\rm w}$. 
 This   is widely  observed and   implies that  linear response and autocorrelation are
again proportional. An  effective temperature can then be defined, albeit in a non-universal 
manner~\cite{Calabrese05} via the Fluctuation Dissipation ratio..

Numerical studies of domain  growth convincingly show that the characteristic linear size
of a thermalized domain increases algebraically in time~\cite{Rieger93, Komori00b}.
In this context, the  $T$ dependence of the growth  exponent  $\lambda$ indicates that the 
energy barrier surmounted at age $t$  grows  logarithmically with $t$. The proposed
explanation  is however  \emph{ad hoc}
and  differs from  the original assumptions of Fisher and Huse~\cite{Fisher88},
 who expected  a power-law scaling.
In record dynamics,  the  same logarithmic scaling  flows   from the proportionality 
between said barrier and the typical size of extremal thermal fluctuations at age $t$. 
 This property leads  directly to the temperature dependence of the exponent $x$
 given in Eq.~\eqref{th_eff_age}. 
Thus, even though  the linear domain size $R(t)$  does not appear 
conspicuously in the present description,   central aging   features 
are clearly associated to properties of   localized  domains.
Firstly, the presence of  non-interacting  domains 
can    be deduced from the statistical properties of energy 
 and linear response fluctuations~\cite{Sibani05,Sibani06b,Christiansen07}, e.g.
it produces  the linear system size dependence of the factor $\alpha$. Secondly, 
 an approximately   hierarchical  energy landscape   is required:  In such  landscape  
 energy fluctuations of record
size are, by construction, necessary  to trigger  any changes of metastable attractor.
Last but not least, the temperature independence of $\alpha$"\cite{Sibani07,Christiansen07} is only possible if
the landscape is invariant to a change of energy scale. 

 In conclusion, a record dynamics description accounts for:  
 \emph{i)}  the linear response phenomenology of aging systems, 
 including the effect of temperature shifts. \emph{ii)}
 The origin of  asymptotic descriptions
in term of   effective temperatures. \emph{iii)}
The interplay of   real space properties, i.e. domains,  and 
 hierarchical  properties associated to the energy landscape of each domain.  
\section{Acknowledgments} Financial support  from the Danish Natural Sciences Research Council
is gratefully acknowledged.  The authors are indebted to 
the Danish Center for Super Computing (DCSC)  for computer time on
the Horseshoe Cluster, where  most of the simulations were carried out. 
%\newpage 
\bibliographystyle{unsrt}
\bibliography{SD-meld}

\begin{thebibliography}{10}

\bibitem{Lundgren86}
L.~Lundgren, P.~Nordblad, and L.~Sandlund.
\newblock Memory behaviour of the spin glass relaxation.
\newblock {\em Europhysics Letters}, 1:529--534, 1986.

\bibitem{Schultze91}
C.~Schultze, K.~H. Hoffmann, and P.~Sibani.
\newblock Aging phenomena in complex systems: A hierarchical model for
  temperature step experiments.
\newblock {\em Europhys. Lett.}, 15:361--366, 1991.

\bibitem{Sibani91}
P.~Sibani and K.H. Hoffmann.
\newblock Relaxation in complex systems : local minima and their exponents.
\newblock {\em Europhys. Lett.}, 16:423--428, 1991.

\bibitem{Hoffmann97}
S.~Schubert K.H.~Hoffmann and P.~Sibani.
\newblock Age reinitialization in spin-glass dynamics and in hierarchical
  relaxation models.
\newblock {\em Europhys. Lett.}, 38:613--618, 1997.

\bibitem{Jonason98}
K.~Jonason, E.~Vincent, J.~Hammann, J.~P. Bouchaud, and P.~Nordblad.
\newblock {M}emory and {C}haos {E}ffects in {S}pin {G}lasses.
\newblock {\em Phys. Rev. Lett.}, 81:3243--3246, 1998.

\bibitem{Komori99}
{T. Komori, H. Yoshino and H. Takayama}.
\newblock {Numerical study on aging dynamics in the 3D Ising spin-glass model.
  I. Energy relaxation and domain coarsening}.
\newblock {\em J. Phys. Soc. Japan}, 68:3387--3393, 1999.

\bibitem{Komori00a}
{T. Komori, H. Yoshino and H. Takayama}.
\newblock {Numerical study on aging dynamics in the 3D Ising spin-glass model.
  II. Quasi-equilibrium regime of spin autocorrelation function}.
\newblock {\em J. Phys. Soc. Japan}, 69:1192--1201, 2000.

\bibitem{Komori00b}
{T. Komori, H. Yoshino and H. Takayama}.
\newblock {Numerical study on aging dynamics in Ising spin-glass models.
  Temperature change protocols }.
\newblock {\em J. Phys. Soc. Japan}, 69:228--237, 2000.

\bibitem{Normand00}
Val\'{e}ry Normand, St\'{e}phane Muller, Jean-Claude Ravey, and Alan Parker.
\newblock Gelation kinetics of gelatin: A master curve and network modeling.
\newblock {\em Macromolecules}, 33:1063--1071, 2000.

\bibitem{Utz00}
Marcel Utz, Pablo~G. Debenedetti, and Frank~H. Stillinger.
\newblock Atomistic simulation of aging and rejuvenation in glasses.
\newblock {\em Phys. Rev. Lett.}, 84:1471--1474, 2000.

\bibitem{Bouchaud01}
Jacques~Hammann Jean-Philippe~Bouchaud, Vincent~Dupuis and Eric Vincent.
\newblock Separation of time and length scales in spin glasses: Temperature as
  a microscope.
\newblock {\em Phys. Rev. B}, 65:024439, 2001.

\bibitem{Nicodemi01}
Mario Nicodemi and Henrik~Jeldtoft Jensen.
\newblock Aging and memory phenomena in magnetic and transport properties of
  vortex matter: a brief review.
\newblock {\em J. Phys A}, 34:8425, 2001.

\bibitem{Bonn02}
{D.Bonn, S. Tanase, B. Abou, H. Tanaka and J. Meunier}.
\newblock Laponite: Aging and shear rejuvenation of a colloidal glass.
\newblock {\em Phys. Rev. Lett.}, 89:015701, 2002.

\bibitem{Berthier02}
Ludovic Berthier and Jean-Philippe Bouchaud.
\newblock Geometrical {A}spects of {A}ging and {R}ejuvenation in the {I}sing
  {S}pin {G}lass: {A} {N}umerical {S}tudy.
\newblock {\em Phys. Rev. B}, 66:054404, 2002.

\bibitem{Takayama02}
{Hajime Takayama and Koji Hukushima}.
\newblock {Numerical Study on Aging Dynamics in 3D Ising Spin-Glass Model. III.
  Cumulative Memory and `Chaos' Effects in the Temperature Shift Protocol.}
\newblock {\em J. Phys. Soc. JPN}, 71:3003--3010, 2002.

\bibitem{Jensen02}
H.~Jeldtoft Jensen and Mario Nicodemi.
\newblock Memory effects in response functions of driven vortex matter.
\newblock {\em Europhys. Lett.}, 57:348, 2002.

\bibitem{Sibani04a}
Paolo Sibani and Henrik~Jeldtoft Jensen.
\newblock How a spin-glass remembers. memory and rejuvenation from
  intermittency data: an analysis of temperature shifts.
\newblock {\em JSTAT}, page P10013, 2004.

\bibitem{Rieger93}
H.~Rieger.
\newblock Non-equilibrium dynamics and aging in the three dimensional {I}sing
  spin-glass model.
\newblock {\em J. Phys. A}, 26:L615--L621, 1993.

\bibitem{Vincent91}
E.~Vincent.
\newblock Slow dynamics in spin glasses and other complex systems.
\newblock In D.~H. Ryan, editor, {\em Recent progress in random magnets}, pages
  209--246. Mc Gill University, 1991.

\bibitem{Dupuis02}
{V. Dupuis, E. Vincent, J.-P. Bouchaud, J. Hammann, A. Ito and H. Aruga
  Katori}.
\newblock Aging, rejuvenation and memory effects in {I}sing and {H}eisenberg
  spin glasses.
\newblock {\em Phys. Rev. B}, 64:174204, 2002.

\bibitem{Andersson92}
J-O. Andersson, J.~Mattsson, and P.~Svedlindh.
\newblock Monte {C}arlo studies of {I}sing spin-glass systems: {A}ging behavior
  and crossover between equilibrium and nonequilibrium dynamics.
\newblock {\em Phys. Rev. B}, 46:8297--8304, 1992.

\bibitem{Sibani07}
{P. Sibani}.
\newblock {Linear response in aging glassy systems, intermittency and the
  Poisson statistics of record fluctuations}.
\newblock {\em Eur. Phys. J. B}, 58:483--491, 2007.

\bibitem{Svedlindh89}
{P. Svedlindh, K. Gunnarsson, P. Nordblad, L. Lundgren, H. Aruga and A. Ito}.
\newblock Equilibrium magnetic fluctuations of a short range ising spin glass.
\newblock {\em Phys. Rev. B}, 40:7162--7166, 1989.

\bibitem{Herisson02}
D.~H{\'{e}}risson and M.~Ocio.
\newblock Fluctuation-dissipation ratio of a spin glass in the aging regime.
\newblock {\em Phys. Rev. Lett.}, 88:257202, 2002.

\bibitem{Buisson03}
{L.~Buisson, L.~Bellon, and S.~Ciliberto}.
\newblock Intermittency in aging.
\newblock {\em J. Phys. Condens. Matter.}, 15:S1163, 2003.

\bibitem{Cugliandolo97}
{Leticia F. Cugliandolo, Jorge Kurchan, and Luca Peliti}.
\newblock Energy flow, partial equilibration, and effective temperature in
  systems with slow dynamics.
\newblock {\em Phys. Rev. E}, 55:3898--3914, 1997.

\bibitem{Castillo03}
{ Horacio E. Castillo, Claudio Chamon, Leticia F. Cugliandolo, Jos{\'{e}} Luis
  Iguain and Malcom P. Kenneth}.
\newblock Spatially heterogeneous ages in glassy systems.
\newblock {\em Phys. Rev. B}, 68:13442, 2003.

\bibitem{Calabrese05}
Pasquale Calabrese and Andrea Gambassi.
\newblock Ageing properties of critical systems.
\newblock {\em J. Phys. A}, 38:R133--R193, 2005.

\bibitem{Sibani05}
P.~Sibani and H.~Jeldtoft Jensen.
\newblock Intermittency, aging and extremal fluctuations.
\newblock {\em Europhys. Lett.}, 69:563--569, 2005.

\bibitem{Sibani06a}
{Paolo Sibani, G.F. Rodriguez and G.G. Kenning}.
\newblock Intermittent quakes and record dynamics in the thermoremanent
  magnetization of a spin-glass.
\newblock {\em Phys. Rev. B}, 74:224407, 2006.

\bibitem{Sibani06b}
{Paolo Sibani}.
\newblock Aging and intermittency in a p-spin model.
\newblock {\em Phys. Rev. E}, 74:031115, 2006.

\bibitem{Christiansen07}
{S. Christiansen and P. Sibani}.
\newblock Linear response and spontaneous fluctuations of aging systems: the
  p-spin model.
\newblock {\em arXiv:0709.1085 [cond-mat.stat-mech]}, 2007.

\bibitem{Crisanti04}
A.~Crisanti and F.~Ritort.
\newblock Intermittency of glassy relaxation and the emergence of a
  non-equilibrium spontaneous measure in the aging regime.
\newblock {\em Europhys. Lett.}, 66:253--259, 2004.

\bibitem{Sibani93}
P.~Sibani, C.~Sch{\"{o}}n, P.~Salamon, and J.-O. Andersson.
\newblock Emergent hierarchical structures in complex system dynamics.
\newblock {\em Europhys. Lett.}, 22:479--485, 1993.

\bibitem{Sibani93a}
P.~Sibani and Peter~B. Littlewood.
\newblock Slow {Dynamics} from {Noise} {A}daptation.
\newblock {\em Phys. Rev. Lett.}, 71:1482--1485, 1993.

\bibitem{Sibani03}
Paolo Sibani and Jesper Dall.
\newblock {Log-Poisson statistics and pure aging in glassy systems.}
\newblock {\em Europhys. Lett.}, 64:8--14, 2003.

\bibitem{Cipelletti00}
Luca Cipelletti, S.~Manley, R.~C. Ball, and D.~A. Weitz.
\newblock Universal aging features in the restructuring of fractal colloidal
  gels.
\newblock {\em Phys. Rev. Lett.}, 84:2275--2278, 2000.

\bibitem{Josserand00}
Christophe Josserand, Alexei~V. Tkachenko, Daniel~M. Mueth, and Heinrich~M.
  Jaeger.
\newblock Memory effects in granular materials.
\newblock {\em Phys. Rev. Lett.}, 85:3632--3635, 2000.

\bibitem{Hannemann05}
A.~Hannemann, J.~C. Sch{\"{o}}n, M.~Jansen, and P.~Sibani.
\newblock {Non-equilibrium dynamics in amorphous {Si$_3$B$_3$N$_7$}}.
\newblock {\em J. Chem. Phys.}, B 109:11770--11776, 2005.

\bibitem{Oliveira05}
{L.P. Oliveira, Henrik Jeldtoft Jensen, Mario Nicodemi and Paolo Sibani}.
\newblock Record dynamics and the observed temperature plateau in the magnetic
  creep rate of type ii superconductors.
\newblock {\em Phys. Rev. B}, 71:104526, 2005.

\bibitem{Anderson04}
{Paul Anderson, Henrik Jeldtoft Jensen, L.P. Oliveira and Paolo Sibani}.
\newblock Evolution in complex systems.
\newblock {\em Complexity}, 10:49--56, 2004.

\bibitem{Lipowski00}
A.~Lipowski and D.~Johnston.
\newblock Cooling-rate effects in a model of glasses.
\newblock {\em Phys. Rev. E}, 61:6375--6382, 2000.

\bibitem{Swift00}
{Michael. R. Swift, Hemant Bokil, Rui D. M. Travasso and Alan J. Bray}.
\newblock Glassy behavior in a ferromagnetic p-spin model.
\newblock {\em Phys. Rev. B}, 62:11494--11498, 2000.

\bibitem{Dall01}
Jesper Dall and Paolo Sibani.
\newblock {Faster} {M}onte {C}arlo simulations at low temperatures. {T}he
  waiting time method.
\newblock {\em Comp. Phys. Comm.}, 141:260--267, 2001.

\bibitem{Sibani06}
Paolo Sibani.
\newblock Mesoscopic fluctuations and intermittency in aging dynamics.
\newblock {\em Europhys. Lett.}, 73:69--75, 2006.

\bibitem{Granberg88}
P.~Granberg, L.~Sandlund, P.~Nordblad, P.~Svedlindh, and L.~Lundgren.
\newblock Observation of a time-dependent spatial correlation length in a
  metallic spin glass.
\newblock {\em Phys. Rev. B}, 38:7097--7100, 1988.

\bibitem{Fisher88}
Daniel~S. Fisher and David~A. Huse.
\newblock Equilibrium behavior of the spin-glass ordered phase.
\newblock {\em Phys. Rev. B}, 38:386--410, 1988.

\end{thebibliography}
\end{document}